\documentclass[fleqn,usenatbib]{mnras}
\usepackage{newtxtext,newtxmath}    
\usepackage[T1]{fontenc}
\usepackage{graphicx}
\graphicspath{{img/}{../Results/NGC2808/fig/paper/}{../Results/NGC2808/fig/final/}}
\usepackage{amsmath}
\DeclareRobustCommand{\VAN}[3]{#2}
\let\VANthebibliography\thebibliography
\def\thebibliography{\DeclareRobustCommand{\VAN}[3]{##3}\VANthebibliography}
\newcommand{\hst}{\textit{HST}}
\newcommand{\jwst}{\textit{JWST}}
\newcommand{\masyr}{mas\,yr$^{-1}$}
\newcommand{\rh}{R_{\rm h}}
\newcommand{\rc}{R_{\rm c}}

\newcommand{\msun}{{\rm M}_{\odot}}
\newcommand{\filt}[1]{\ensuremath{m_{\mathrm{F#1W}}}}
\NewDocumentCommand{\cmd}{>{\SplitArgument{2}{,}}m}{%
  \cmdfilters#1%
}

\newcommand{\cmdfilters}[3]{%
  \ensuremath{
    \filt{#1}\;\mathrm{vs.}\;\filt{#2}-\filt{#3}
  }%
}

\title[Internal dynamics of NGC\,2808]{Globular Clusters in the Time of JWST. II. Looks like the second, moves like the first: the syrphid-like intermediate population of NGC\,2808}
\author[G. Cordoni et al.]{
G.\,Cordoni,$^{1}$\thanks{E-mail: giacomo.cordoni@anu.edu.au}
A.\,P.\,Milone,$^{2,3}$
E.\,Dondoglio,$^{4, 3}$
A.\,F.\,Marino,$^{3}$
F.\,D'Antona,$^{5}$
A.\,L.\,Varri,$^{6,7}$
\newauthor
V.\,Altomonte,$^{2}$
P.\,Bianchini,$^{8}$
E.\,Bortolan,$^{2}$
L.\,Casagrande,$^{1}$
Y.\,Cavecchi,$^{9, 10}$
G.\,S. Da Costa,$^{1}$
\newauthor
F.\,Dell'Agli,$^{5}$
H.\,Jerjen,$^{1}$
E.\,P.\,Lagioia,$^{11}$
M.\,V.\,Legnardi,$^{2}$
C.\,Li,$^{12,13}$
A.\,Mastrobuono-Battisti,$^{2,3,14}$
\newauthor
F.\,Muratore,$^{2}$
A.\,Renzini,$^{3}$
M.\,Tailo,$^{15}$
P.\,Ventura,$^{5}$
and T.\,Ziliotto$^{16}$
\\
$^{1}$Research School of Astronomy \& Astrophysics, Australian National University, Canberra, ACT 2611, Australia\\
$^{2}$Dipartimento di Fisica e Astronomia ``Galileo Galilei'', Universit\`a di Padova, Vicolo dell'Osservatorio 3, 35122 Padova, Italy\\
$^{3}$Istituto Nazionale di Astrofisica - Osservatorio Astronomico di Padova, Vicolo dell'Osservatorio 5, 35122 Padova, Italy\\
$^{4}$Physics Department, American University of Sharjah, P.O. Box 26666, Sharjah, UAE\\
$^{5}$Istituto Nazionale di Astrofisica - Osservatorio Astronomico di Roma, Via Frascati 33, 00077 Monte Porzio Catone, Italy\\
$^{6}$School of Mathematics and Maxwell Institute for Mathematical Sciences, University of Edinburgh, Kings Buildings, Edinburgh EH9 3FD, UK\\
$^{7}$Institute for Astronomy, University of Edinburgh, Royal Observatory, Blackford Hill, Edinburgh EH9 3HJ, UK\\
$^{8}$Universit\'e de Strasbourg, CNRS, Observatoire astronomique de Strasbourg, UMR 7550, 67000 Strasbourg, France\\
$^{9}$Departamento de Astrof\'{\i}sica, Universidad de La Laguna, 38206 San Crist\'obal de La Laguna, Tenerife, Spain\\
$^{10}$Instituto de Astrof\'{\i}sica de Canarias, E-38205 San Crist\'obal de La Laguna, Tenerife, Spain\\
$^{11}$South-Western Institute for Astronomy Research, Yunnan University, Kunming 650500, P. R. China\\
$^{12}$School of Physics and Astronomy, Sun Yat-sen University, Daxue Road, Zhuhai 519082, P. R. China\\
$^{13}$CSST Science Center for the Guangdong-Hong Kong-Macau Greater Bay Area, Zhuhai 519082, P. R. China\\
$^{14}$Dipartimento di Tecnica e Gestione dei Sistemi Industriali, Universit\`a degli Studi di Padova, Stradella S. Nicola 3, 36100 Vicenza, Italy\\
$^{15}$Dipartimento di Fisica e Astronomia ``Augusto Righi'', Universit\`a di Bologna, Via Gobetti 93/2, 40129 Bologna, Italy\\
$^{16}$Space Telescope Science Institute, 3700 San Martin Drive, Baltimore, MD 21218, USA
}
\date{Accepted XXX. Received YYY; in original form ZZZ}
\pubyear{2026}
\begin{document}
\label{firstpage}
\pagerange{\pageref{firstpage}--\pageref{lastpage}}
\maketitle

\begin{abstract}

The present-day kinematics of multiple populations in globular clusters can retain information about their initial conditions and can therefore help trace how the different populations formed. We combine \jwst\ and \hst\ photometry and proper motions to study the internal dynamics of four populations in NGC\,2808 out to about eight half-light radii: the first population (1P) and three second-population groups with increasing light-element processing (2Pa, 2Pb, and 2Pc), providing the first detailed kinematic characterisation of the chemically intermediate 2Pa. We find significant kinematic differences between the populations. While all four have similar sky-projected radial velocity dispersions, 2Pb and 2Pc have lower tangential dispersions than 1P and are strongly radially anisotropic, whereas 2Pa shares the only mildly anisotropic kinematics of 1P. Therefore, chemically distinct 2P groups have different dynamical properties, suggesting that they formed under different conditions and/or experienced a different subsequent dynamical evolution. In the outer \jwst\ field, the velocity dispersion depends only weakly on stellar mass, with hints of inverted equipartition. The difference between radial and tangential equipartition is similar for 1P and 2Pa and decreases towards 2Pc, a trend that remains tentative given the uncertainties. Our results show that 2P is not a dynamically homogeneous group, and that the kinematics of its individual groups provide a key constraint for multiple population formation scenarios.
\end{abstract}

\section{Introduction}
\label{sec:intro}
Photometric and spectroscopic studies have established that nearly all Galactic globular clusters (GCs) host multiple stellar populations (MPs), with chemical differences observed from the main sequence (MS) to the late stages of stellar evolution. First-population (1P) stars have light-element abundances similar to those of field stars at the same metallicity, whereas second-population (2P) stars are enriched in helium, nitrogen, sodium, and aluminium, and depleted in carbon, oxygen, and magnesium \citep{gratton2012, piotto2015, renzini2015, milone2017, bastianlardo2018, milonemarino2022}. In several clusters, such as NGC\,2808, the enriched component contains distinct groups with different degrees of light-element processing \citep[see e.g.][]{marino2019, zennaro2019}.

Several scenarios have been proposed to explain these abundance patterns, but none reproduces all the observational constraints \citep[see reviews by][]{renzini2015, bastianlardo2018, milonemarino2022}. Candidate sources of processed material include asymptotic giant branch stars (AGBs), fast-rotating massive stars, massive interacting binaries (MIBs), and supermassive or extremely massive stars (SMSs and EMSs) \citep{ventura2001, decressin2007, dercole2008, demink2009, denissenkov2014, gieles2018, gieles2025, renzini2022, bekki2026}. These models differ in the source of the ejecta and how it is incorporated into newly forming and/or low-mass stars. In particular, in scenarios where the gas collects near the cluster centre before forming a new stellar generation, the 2P begins in a compact subsystem embedded in the more extended 1P. This, in turn, translates into different initial spatial and dynamical properties of 1P and 2P stars.

Dynamical evolution can subsequently alter these initial distributions. Two-body relaxation mixes the populations most rapidly in the centre, while differences can survive for longer in the outer regions. In simulations, an initially more concentrated 2P evolves towards radial anisotropy in the outer regions, while the more extended 1P remains closer to isotropy \citep{mastrobuono-battisti2013, mastrobuono-battisti2016, henault-brunet2015, tiongco2019, vesperini2021}. The radial and tangential velocity-dispersion profiles can therefore retain evidence of MPs' initial spatial configurations, particularly where cluster relaxation is incomplete.

Over the past decade, Hubble Space Telescope (\hst) proper motions have allowed these predictions to be tested in Galactic GCs, revealing stronger radial anisotropy among enriched stars in several clusters \citep{richer2013, bellini2015, bellini2018, milone2018, libralato2023}. More recently, ground-based photometry combined with \textit{Gaia} and \hst\ proper motions have extended the analysis to larger cluster samples and wider fields \citep{cordoni2020a, cordoni2020b, dalessandro2024, leitinger2025, cordoni2025a}. These studies find that dynamically younger clusters, where the relaxation process is still incomplete, exhibit a clear dynamical distinction between 1P and 2P stars. Specifically, 2P stars display a radially anisotropic motion in the outer regions (e.g., outside one half-light radius, $\rh$), whereas 1P stars are isotropic or show mild tangential anisotropy. However, because of the lower quality of ground-based observations, which only allows the separation of MPs into 1P and 2P, little is known about the chmically distinct 2P sub-groups.

In this context, NGC\,2808 represents an ideal case. The analysis of the chromosome map (ChM) reveals that, in addition to an extended 1P hosting stars with different metallicities \citep{legnardi2022, lardo2023}, this cluster hosts at least three groups of enriched stars\footnote{We use the term enriched stars to indicate 2P stars, even though they are depleted in some elements, such as oxygen and carbon.}, labelled C, D, E populations in \citet{milone2015}.

It is now known that its split MS contains helium-rich populations reaching $Y\sim 0.38$, and ultraviolet (UV) photometry resolves at least five groups along the MS and red giant branch (RGB) \citep{dantona2005, piotto2007, milone2012, milone2015}.  Spectroscopy also reveals variations in helium and in various light elements, including C, N, O, Na, Mg, Al, Si, and K \citep{carretta2006a, marino2011, marino2014b, carlos2023}. 

From the dynamical perspectives, the kinematics of the individual populations of NGC\,2808 were first investigated with \hst\ by \citet{bellini2015}, who found that the more enriched populations D and E are more radially anisotropic than A, B, and C, although their analysis was limited to the central regions of the cluster. \citet{griggio2025} extended the analysis to larger radii, finding radial anisotropy for D and E and nearly isotropic kinematics for the red MS (rMS). However, as the rMS combines populations A, B, and C, the kinematics of the chemically intermediate population C could not be separated from those of the 1P.

This paper is part of our series Globular Clusters in the Time of JWST, which aims to exploit the capabilities of the James Webb Space Telescope (\jwst) to investigate Galactic GCs, with particular emphasis on their multiple populations. In the first paper of the series \citep{milone2026}, we presented the survey design and initial results for 11 GCs, including the detection and characterisation of MPs among low-mass stars.
In this work, we combine \jwst\ observations (GO-8960) with archival \hst\ observations to extend previous studies and investigate the internal dynamics of 1P stars of NGC\,2808 (corresponding to the A+B groups) and of the three main groups of 2P stars, 2Pa, 2Pb, and 2Pc, which correspond to groups C, D, and E, respectively. This is the first time the dynamical properties of population 2Pa (C) are studied at such a level of detail and wide field of view (from the cluster centre out to $\sim8$ half-light radii, $\rh=0.80$\,arcmin; \citealt[][2010 edition]{harris1996}). We then use these results to discuss the formation history of NGC\,2808 and of its chemically intermediate 2Pa. Our analysis includes both the sky-projected radial profile of velocity dispersions and the dependence of velocity dispersion on stellar mass, i.e. energy equipartition. 

This work is organised as follows: Secs.~\ref{sec:data} and~\ref{sec:mps} describe the observations and the selection of MPs, while Sec.~\ref{sec:methods} presents the dynamical analysis. Results are presented and discussed in Secs.~\ref{sec:results} and~\ref{sec:discussion and conclusions}, respectively, while Sec.~\ref{sec:summary} summarises the results. 

\section{Data and data reduction}
\label{sec:data}

\begin{figure}
    \centering
    \includegraphics[width=0.45\textwidth]{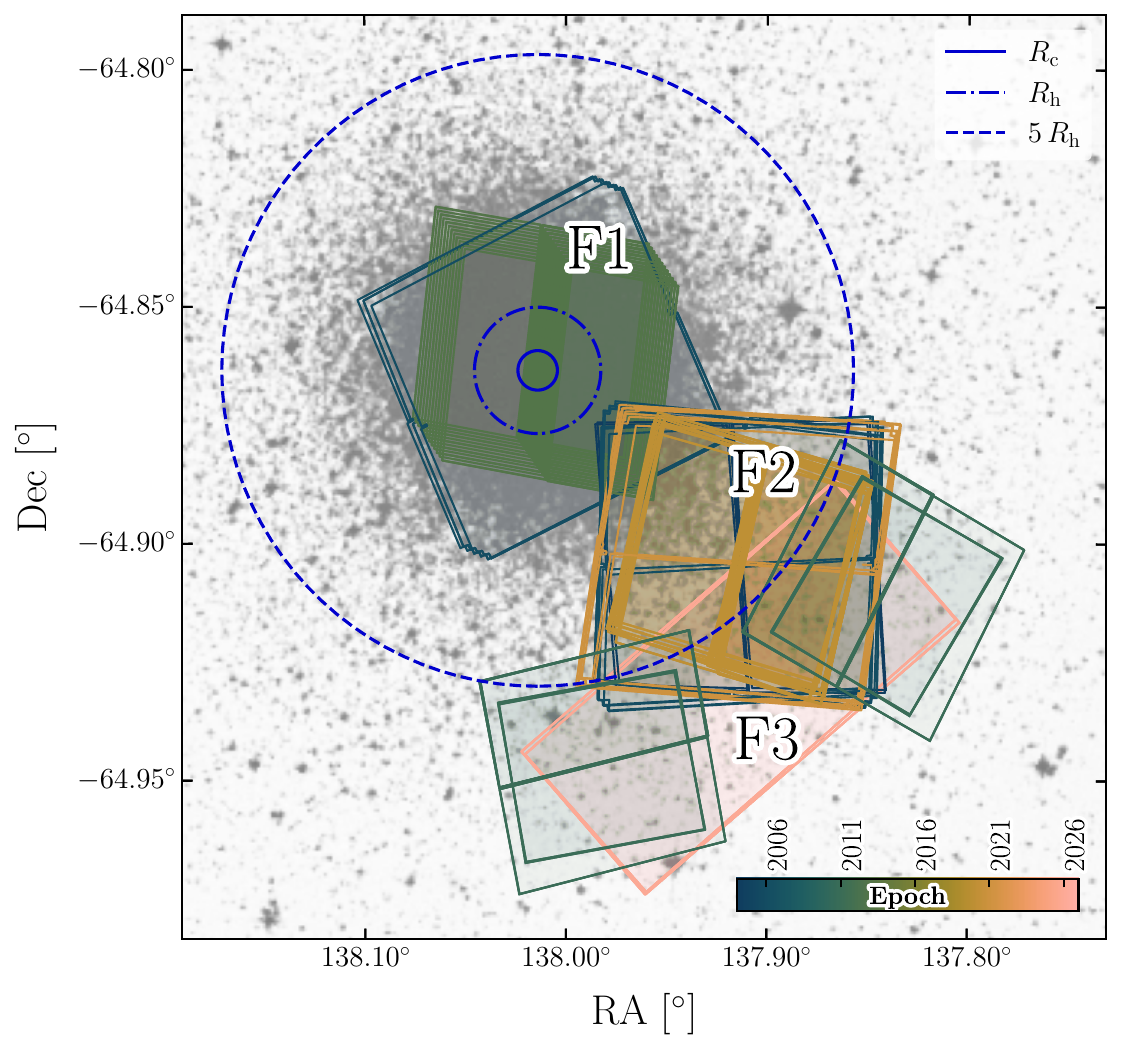}
    \caption{Footprints of the \hst\ (F1 and F2) and \jwst\ (F3) observations used in this work to identify MPs, together with the two additional \hst\ fields used only to derive proper motions in F3, overlaid on a DSS2 blue image of NGC\,2808 (greyscale, darker for brighter regions). Colours indicate the observing epoch, while blue circles mark the core ($\rc$), half-light ($\rh$) and five times the half-light ($5\,\rh$) radii for reference.}
    \label{fig:footprints}
\end{figure}

To investigate the internal dynamics of MPs in NGC\,2808, we use the \jwst\ observations presented in \citet{milone2026}. The dataset combines \jwst\ Near-Infrared Camera (NIRCam) images in F115W, F200W, F277W and F444W recently acquired as part of program GO-8960 (PI Milone, A.~P. and Marino, A.~F.) and archival \hst\ ACS/WFC and WFC3/UVIS images, as shown in the footprints in Fig.~\ref{fig:footprints}.  The \hst\ data serve two purposes. Fields F1 and F2 include ultraviolet and optical observations suited to identify MPs, and are used for both photometry and proper motions. Two additional \hst\ fields overlap with the \jwst\ field F3 and are used exclusively as earlier astrometric epochs to measure proper motions across the whole F3 field. Together, the three fields cover projected distances from the cluster centre out to about $\simeq6.4$~arcmin ($\simeq8~\rh$): F1 samples the central regions out to $\sim2.5~\rh$, F2 covers $\sim2$--$6\,\rh$, and F3 extends from $\sim4.5$ to $\sim8~\rh$. Details of the \jwst\ (F3) and \hst\ (F2) observations are available at \url{https://doi.org/10.17909/ztzd-kn19}, while for F1 we refer to \citet{libralato2022}.

\begin{figure*}
    \centering
    \includegraphics[width=0.99\textwidth]{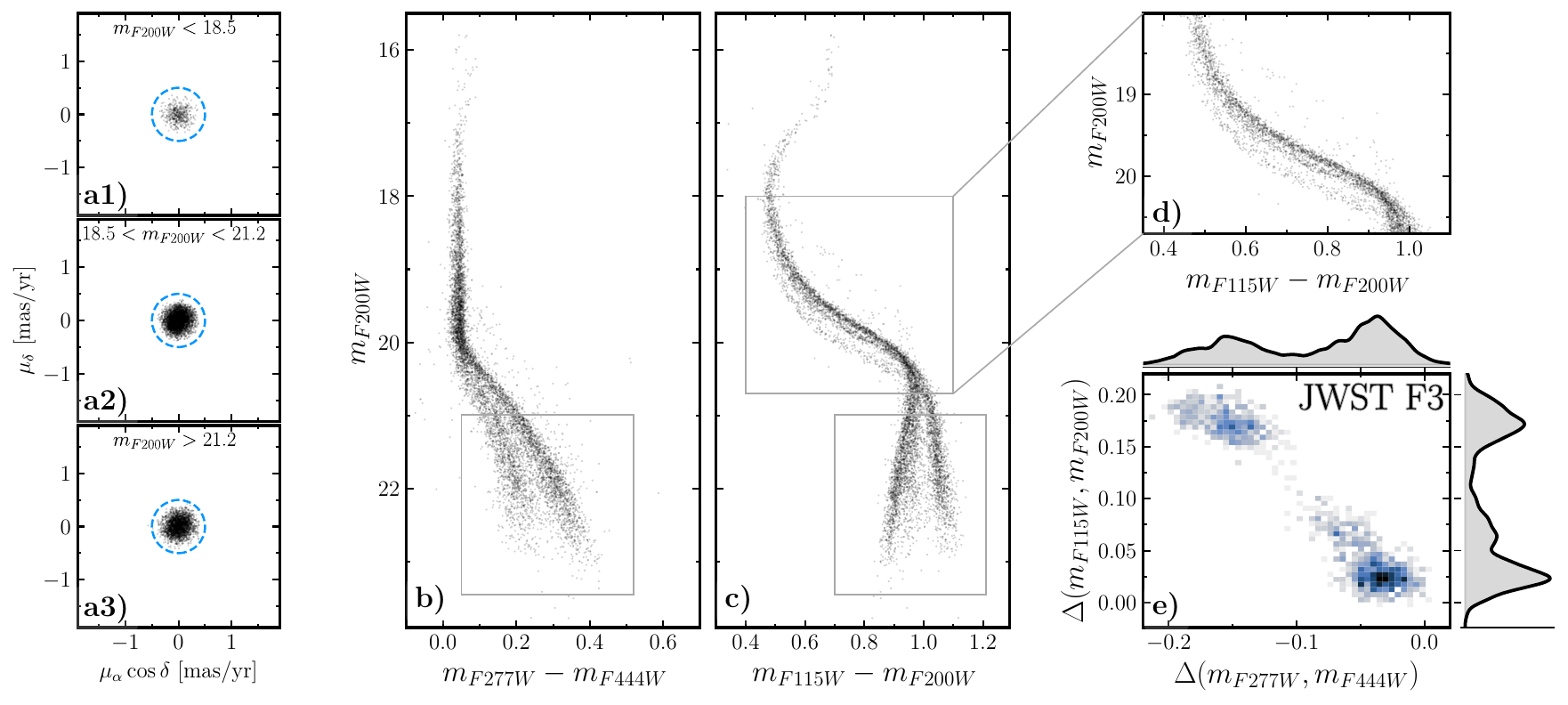}
    \caption{Photometry and proper motions of \jwst\ observations for field F3. Panels a1-a3) illustrate the cluster members (black) in the vector-point diagrams for three magnitude intervals, with the dashed circle of radius 0.5\,\masyr, centred on the cluster mean motion, shown for reference only. Panels b) and c) show the \jwst\ \cmd{200, 277, 444} and \cmd{200, 115, 200} colour-magnitude diagrams (CMDs), while panel d) displays a zoom-in around bright MS stars in the latter CMD. Panel e) represents the \jwst\ ChM derived from the CMDs in panels b) and c) for stars in the grey boxes, with KDE distributions represented with black lines in the marginal plots.}
    \label{fig:photometry jwst}
\end{figure*}

The ultraviolet \hst\ colours separate multiple populations among upper-MS and evolved populations like red-giant branch (RGB) stars, while the infrared \jwst\ colours identify MPs among MS and M-dwarfs. In the following, we provide a brief description of \jwst\ astro-photometric data reduction, but refer to \citet{milone2026} for a full discussion of the adopted methodology. The resulting \jwst\ photometry is shown in Fig.~\ref{fig:photometry jwst}.

\subsection{JWST photometry}
\label{subsec:jwst_reduction}
Data reduction of \jwst\ images has been carried out as described in \citet{milone2026}. Here we provide a brief description, but refer to the paper for a complete and detailed discussion. Briefly, we measured stellar positions and fluxes using the KS2 package \citep{sabbi2016, bellini2017a, milone2023a}, which builds on the approach introduced by \citet{anderson2008a}. The software analyses all exposures simultaneously and tailors the measurement procedure to the brightness of each source. Bright, unsaturated stars were measured by fitting each exposure with a spatially varying effective point-spread function \citep[ePSF;][]{anderson2000a}. In contrast, faint stars in crowded regions were measured using weighted-aperture photometry after the light from neighbouring stars had been subtracted. For all sources, the measurements obtained in the individual exposures were combined into final magnitudes and positions.  Instrumental magnitudes were placed on the Vega system using the STScI encircled-energy corrections and zero points for NIRCam filters, while pixel-area and geometric-distortion corrections were applied using the appropriate calibration solutions \citep{anderson2006a, bellini2009a, bellini2011a, anderson2022a}.

KS2 output diagnostics were used to reject poorly measured stars \citep[see e.g.,][]{nardiello2018} and retain only sources with the best photometry and astrometry, resulting in a high-quality sample required to identify MPs in different CMDs.

Relative proper motions in F2 and F3 were computed by comparing stellar positions measured at different epochs with \jwst\ and \hst\, following established high-precision astrometric techniques \citep[e.g.][]{anderson2003, piotto2012a, bellini2014a, libralato2022, milone2023a, ziliotto2025, ziliotto2026}, and as discussed in \citet{milone2026}. The catalogues from all filters and epochs were aligned to a common reference frame defined by the F200W NIRCam images. The alignment was performed with six-parameter linear transformations based on bright cluster members, whose membership was refined iteratively from the CMD and proper-motion diagrams. Local transformations, computed from nearby reference stars and excluding the target star, were used to account for residual small-scale distortions. Finally, proper motions were derived from weighted linear fits of the transformed positions versus epoch.

The resulting proper motions and photometric diagrams are presented in Fig.~\ref{fig:photometry jwst}. Specifically, cluster members, identified as in \citet[][see their Fig.~8]{milone2026}, are displayed in panels a1--a3 as black points for different magnitude intervals, and the blue dashed circles indicate $0.5$\,\masyr. The \cmd{200, 277, 444} and \cmd{200, 115, 200} CMDs (panels b and c) have been used to build the ChM of faint M-dwarf stars displayed in panel e. For the analysis of the ChM, we focus on the stars in the grey boxes, where MPs are most evident (see Sec.~\ref{sec:mps} for a description of the selection procedure). The marginal plots display kernel density estimates \citep[KDEs;][]{scikit-learn} along the $x$ and $y$ axes. Panel d shows a zoom-in of the \cmd{200, 115, 200} CMD around bright MS stars, used in Sec.~\ref{sec:mps} to identify MPs. The ChM of cluster members has been constructed as in \citet{milone2026}.

\subsection{HST photometry}
\label{subsec:hst}
We use two sets of \hst\ observations. Fields F1 and F2 cover ultraviolet and optical filters and are used to identify MPs. Field F1 samples the inner cluster, with photometry from \citet{milone2015} and relative proper motions from \citet{libralato2022}. These catalogues identify populations among bright-MS and evolved stars, and we refer to \citet{libralato2022} for details about how proper motions were computed.

Field F2 overlaps with the \jwst\ field F3 and was re-reduced for this work. We processed ACS/WFC and WFC3/UVIS images with KS2 and placed the positions and photometry on the reference system of the \jwst\ catalogues. In the overlapping region (top right corner of \jwst\ F3 field), the \hst\ and \jwst\ photometry was combined, while proper motions are those derived including \jwst\ observations.

In addition, two \hst\ fields, whose observations are also listed at the DOI above, cover most of the \jwst\ field F3 (Fig.~\ref{fig:footprints}). As they lack ultraviolet filters, these observations are not used to identify MPs, but only as earlier astrometric epochs to derive proper motions over the whole F3 field, beyond its overlap with F2.

\begin{figure}
    \centering
    \includegraphics[width=0.4\textwidth]{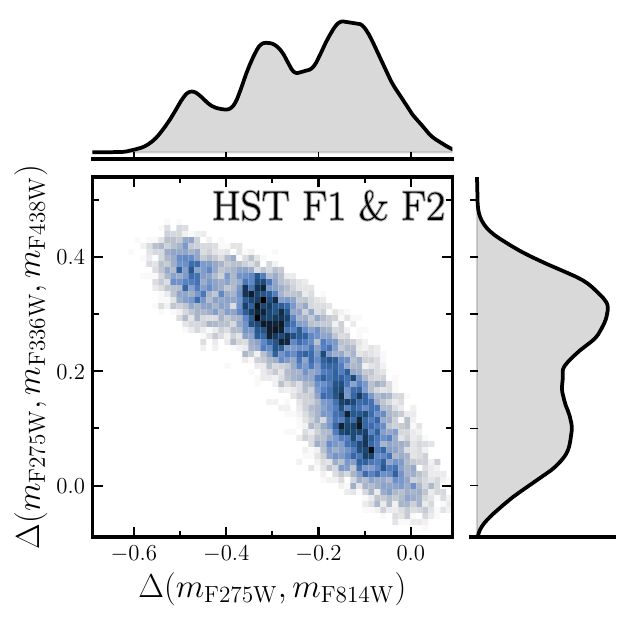}
    \caption{Combined \hst\ ChM for RGB and bright MS stars of field F1 and F2. ChM is built from \filt{275}, \filt{336}, \filt{438}, and \filt{814} as in \citet{milone2015}. Bright MS stars extend down to about $0.55\,\msun$.}
    \label{fig:hst chm}
\end{figure}

\hst\ data for F1 and F2 include observations in  F275W, F336W, F438W and F814W, which have been widely used to identify and study MPs with light-element abundance variation. To this end, we followed the procedure described in \citet{milone2015} to build ChMs for both photometric datasets, spanning the RGB and the upper MS. All ChMs have been combined onto a common reference frame and the combined ChM is displayed in Fig.~\ref{fig:hst chm}. The combination is performed only for a visual representation, while the selection discussed in Sec.~\ref{sec:mps} is performed on the individual ChMs, separating RGB and MS in both F1 and F2. KDEs of $x$ and $y$ are shown in the marginal plots.

\section{Selection of multiple populations}
\label{sec:mps}
In this section, we first summarise the populations of NGC\,2808 and the adopted labels, and then describe their selection on the \hst\ ChMs of F1 and F2 and on the \jwst\ ChM and bright MS of F3. 
NGC\,2808 hosts five main stellar populations, originally labelled A--E in the ChM analysis by \citet{milone2015} and \citet{bellini2015}. Populations A and B correspond to the first-population stars, whereas populations C, D, and E are chemically enriched second-population stars. These latter populations exhibit increasing light-element processing, with the C--E populations showing increasing helium abundance and decreasing oxygen abundance. In this work, we refer to populations C, D, and E as 2Pa, 2Pb, and 2Pc, respectively. The correspondence with the labels adopted in previous works is summarised in Table~\ref{tab:populations}. Because populations A and B can be separated only among RGB stars and not among MS and M-dwarfs, we treat them as a single 1P population. This labelling is used throughout this work.

\begin{table}
\centering
\begin{tabular}{cccc}
\hline
\citet{milone2015} & \citet{piotto2007} & This work & Helium \\
\citet{bellini2015} &  &  & $Y$ \\
\hline
A, B & rMS & 1P  & 0.25 \\
C    & rMS & 2Pa & 0.25 \\
D    & mMS & 2Pb & 0.32 \\
E    & bMS & 2Pc & 0.38 \\
\hline
\end{tabular}
\caption{Labels of the populations of NGC\,2808 in previous works and in this work, together with the helium abundances adopted in Sec.~\ref{subsec:masses}.}
\label{tab:populations}
\end{table}

To identify the populations, we use ChMs constructed from verticalised CMDs as described in \citet{milone2015, milone2017} for \hst\ and in \citet{milone2026} for \jwst\ data. This approach is widely used in the literature as the most effective way to identify chemically distinct populations in GCs. \hst\ ultraviolet colours (F275W, F336W, F438W) respond to helium, nitrogen, and oxygen variations and distinguish the four populations among RGB and upper-MS stars \citep{milone2015, milone2017}. Among M-dwarfs, \jwst\ infrared colours are mainly sensitive to oxygen through molecular absorption in their cool atmospheres. Indeed, previous \jwst\ studies have shown that the M-dwarf ChM separates the populations consistently with the \hst\ classification of RGB and upper-MS stars \citep{milone2023b, ziliotto2023, ziliotto2025, ziliotto2026}, and NIRSpec spectroscopy has confirmed that the groups identified on the M-dwarf ChM differ in oxygen abundance \citep{marino2024a, marino2024b}.

For F1 and F2, we constructed \hst\ ChMs from $\Delta(\filt{275}, \filt{814})$ and $\Delta(\filt{275}, \filt{336}, \filt{438})$, which are the verticalised $\filt{275}-\filt{814}$ colour and pseudo-colour $C_{\rm F275W,F336W,F438W}=(\filt{275}-\filt{336})-(\filt{336}-\filt{438})$, used as the $x$ and $y$ axes of the ChM, respectively. We then selected the populations as in \citet{milone2015}, independently for RGB and MS stars in each field. The populations were identified among RGB stars with $15.0 \lesssim \filt{814} \lesssim 18.3$ in F1 and $14.5 \lesssim \filt{814} \lesssim 17.8$ in F2, and among upper-MS stars with $19.6 \leq \filt{814} \leq 20.7$ in F1 and $20.0 \leq \filt{814} \leq 20.5$ in F2. A combined ChM is shown in Fig.~\ref{fig:hst chm}.

For the M-dwarfs in F3 with $20.7 \lesssim \filt{200} \lesssim 22.6$ (grey boxes in Fig.~\ref{fig:photometry jwst}b and c), we constructed the \jwst\ ChM from $\Delta(\filt{277}, \filt{444})$ and $\Delta(\filt{115}, \filt{200})$, the verticalised $\filt{277}-\filt{444}$ and $\filt{115}-\filt{200}$ colours. These colours separate the oxygen-rich 1P from the oxygen-depleted 2Pa, 2Pb, and 2Pc. The boundaries between the populations were determined from the $x$ and $y$ KDE distributions, and the four groups visible in Fig.~\ref{fig:photometry jwst}e, in order of increasing $\Delta(\filt{115}, \filt{200})$, are identified as 1P, 2Pa, 2Pb, and 2Pc. The selection is shown in Fig.~\ref{fig:chm_selection}a, where 1P, 2Pa, 2Pb, and 2Pc are shown in red, orange, blue, and azure, respectively.

To check the effect of MPs selection on the dynamical analysis carried out, we tested the effect of the adopted boundaries by repeating the analysis with Gaussian mixture models \citep[GMM;][]{scikit-learn}
and after removing stars close to the population dividing lines. Such tests are displayed in Fig.~\ref{fig:chm_selection}b and c. We find that the results presented and discussed in the next sections remain consistent in all cases (Appendix~\ref{app:chm_selection}).

While the ChM of M-dwarf allows the selection of four main stellar populations with different oxygen abundance, above the MS knee (corresponding to $\filt{200}\sim 20.5$) such distinction is not possible, as 1P and 2Pa overlap forming the rMS. Above the knee, the three sequences mainly reflect the different helium abundances of the populations, whereas below the knee the M-dwarf colours are dominated by oxygen through molecular absorption. To select these three groups, we exploited the \cmd{200, 115, 200} CMD above the MS knee, shown in Fig.~\ref{fig:photometry jwst}d. Specifically, we computed the three fiducial lines displayed in Fig.~\ref{fig:upperms_gmm} by dividing the CMD into 0.5-magnitude-wide bins and determining the median colour of each bin. The CMD has then been verticalised using a three-fiducial line verticalisation, and the result of such verticalisation is represented in Fig.~\ref{fig:upperms_gmm}b (we refer to Appendix~\ref{app:upperms} for a description of the verticalisation procedure). Finally, we used GMM on three three magnitude bins indicated by the horizontal lines to identify rMS, 2Pb and 2Pc stars (burgundy, blue and azure points, respectively). We retained rMS and 2Pb stars with $19.0 \leq \filt{200} \leq 20.2$, while 2Pc stars, which are clearly separated over a wider magnitude range, are retained over the whole $18.5 \leq \filt{200} \leq 20.5$ interval.

\subsection{Stellar masses}
\label{subsec:masses}
We estimated stellar masses by interpolating BaSTI [$\alpha$/Fe]\,$=+0.4$ isochrones \citep{hidalgo2018, pietrinferni2021} with $[{\rm Fe/H}]=-1.05$ \citep{horta2020} and an age of 11\,Gyr \citep{vandenberg2013}. We adopted a primordial helium abundance, $Y=0.25$ for 1P and 2Pa (corresponding to the rMS population), and $Y=0.32$ and $0.38$ for 2Pb and 2Pc, respectively \citep{marino2014b}.\footnote{Below the MS knee $(\filt{200}\sim20.5)$, the colours of M-dwarfs are mainly sensitive to oxygen through molecular absorption, and isochrones that differ only in helium do not reproduce the observed M-dwarf sequences. Unlike \citet{ventura2026}, we did not compute dedicated isochrones with oxygen-dependent atmospheres, and we derived the M-dwarf masses from the helium-dependent isochrones alone. The resulting mass differences at fixed magnitude are expected to be small and, given the weak mass dependence of the velocity dispersion, not to affect our equipartition results.}

\begin{figure}
    \centering
    \includegraphics[width=0.45\textwidth]{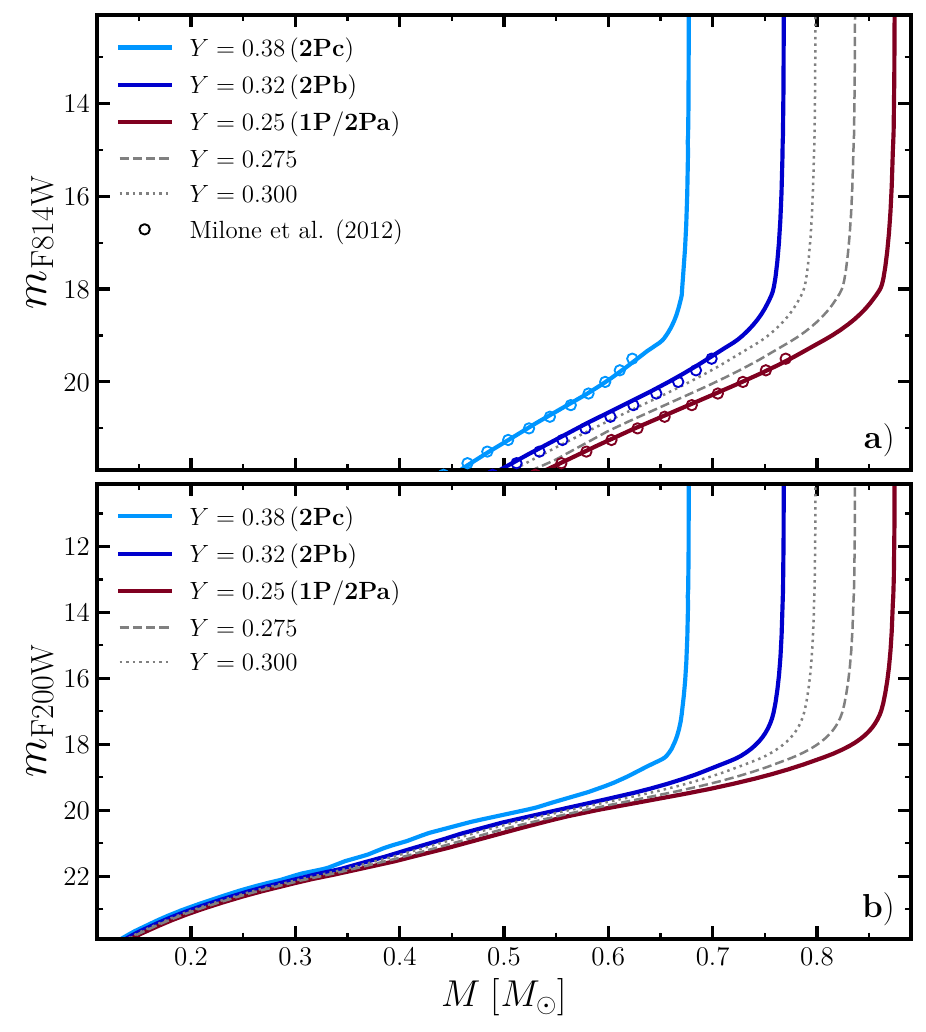}
    \caption{Luminosity--stellar mass relations in \hst\ $\filt{814}$ (panel a) and \jwst\ $\filt{200}$ (panel b) for the three adopted helium abundances, $Y=0.25$ (1P and 2Pa), $0.32$ (2Pb), and $0.38$ (2Pc), derived from BaSTI isochrones with $[{\rm Fe/H}]=-1.05$ and [$\alpha$/Fe]\,$=+0.4$. The $Y=0.32$ and $0.38$ relations are linearly extrapolated in helium from the $Y=0.25$ and $0.30$ isochrones. The original BaSTI isochrones with $Y=0.275$ and $0.30$ are shown as grey dashed and dash-dotted lines, respectively. Open circles in panel a show the masses derived by \citet{milone2012} for the rMS, 2Pb, and 2Pc, coloured as the corresponding isochrones.}
    \label{fig:isochrones}
\end{figure}

As available BaSTI isochrones extend only to $Y=0.30$, we derived the luminosity-stellar mass relations for 2Pb and 2Pc by linear extrapolation in helium, at fixed magnitude, from the $Y=0.25$ and $0.30$ isochrones. Figure~\ref{fig:isochrones} shows the inferred luminosity--mass relations for the \hst\ $\filt{814}$ (panel a) and \jwst\ $\filt{200}$ (panel b) photometry, together with the original BaSTI isochrones with $Y=0.275$ and $0.30$. At fixed magnitude, the helium-rich 2Pb and 2Pc stars have lower masses than 1P and 2Pa stars. In panel a, the masses derived by \citet{milone2012} for the rMS, 2Pb, and 2Pc are shown for comparison. In the subsequent dynamical analysis, we only retain stars with proper-motion uncertainties below 0.1\,\masyr\ in both components, corresponding approximately to stars with $M\gtrsim0.25\,\msun$. The final sample includes about 22\,000 stars of the four populations.

\section{Internal dynamics}
\label{sec:methods}
In this section, we describe the methodology used to investigate the internal dynamics of each population using the sky-projected radial and tangential proper-motion components of each star. We first fit the velocity dispersion and anisotropy profiles as functions of projected distance from the cluster centre, then fit the mass dependence of velocity dispersion, i.e. the energy equipartition. In the following, the internal dynamics is studied by fitting analytical velocity dispersion and anisotropy profiles, similar to \citet{dalessandro2024, ziliotto2025, ziliotto2026}, whereas binned estimates are only used as a visual comparison and check. 

For each star, we computed the sky-projected radial and tangential proper-motion components, $\mu_{\rm R}$ and $\mu_{\rm T}$, by rotating the proper motions along right ascension and declination, $(\mu_{\alpha^*},\mu_\delta)$, into the radial and tangential directions \citep[see e.g., definitions in][]{vasiliev2019a, cordoni2025a}. With our convention, positive $\mu_{\rm R}$ indicates expansion, and positive $\mu_{\rm T}$ corresponds to motion from North towards East, i.e. counter-clockwise on the sky.

\subsection{Likelihood and fitting procedure}
\label{subsec:likelihood}
We model each proper-motion component with a Gaussian distribution, so that the log-likelihood can be written as
\begin{equation}
\begin{aligned}
\ln\mathcal{L}={}&-\frac{1}{2}\sum_{i=1}^{N}\sum_{X\in\{{\rm R,T}\}}
\left[\ln\left(2\pi S_{X,i}^2\right)
+\frac{(\mu_{X,i}-\overline{\mu}_X)^2}{S_{X,i}^2}\right],\\
S_{X,i}^2={}&\sigma_X(R_i)^2+\delta_{X,i}^2,
\end{aligned}
\label{eq:likelihood}
\end{equation}
where $\mu_{X,i}$ and $\delta_{X,i}$ are the measured motion and its uncertainty, and $\overline{\mu}_X$ is the model mean. The variable $R_i$ is the projected radius for the radial profiles, while for energy equipartition it is represented by the stellar mass $m_i$. $\sigma_X(R_i)$ here represents the radial (R) or tangential (T) component of the velocity dispersion profile as described by the analytical models in Secs.~\ref{subsec:profiles} and~\ref{subsec:equipartition}.

% We sample the posterior distribution with the python package \textsc{nautilus} \citep{nautilus}, adopting the same broad priors for all populations within each model, and report the posterior medians and $16-84^\mathrm{th}$ percentile intervals. 
For both the radial profiles and energy equipartition, we tested models with zero mean motions and with constant mean motions of the two components as free parameters, obtaining consistent dispersions. The adopted priors are described in Appendix~\ref{app:bestfit}.

To derive the binned profiles we divided the stars of each field into radial bins with the same number of stars, or in mass for the equipartition analysis. Within each bin, we fitted constant mean motions and dispersions of the two proper-motion components with the same likelihood of Eq.~\eqref{eq:likelihood}, sampled with \textsc{dynesty} \citep{dynesty}, and adopted the posterior median and $16-84^\mathrm{th}$ percentiles as the binned estimate and its uncertainty. Each bin is plotted at the median radius (or mass) of its stars, with horizontal bars spanning the bin edges (see e.g. Fig.~\ref{fig:profiles}). The binned and unbinned profiles are fitted independently to the same catalogue and are shown together only for visual comparison.

\subsection{Velocity-dispersion and anisotropy profiles}
\label{subsec:profiles}
We describe the radial velocity dispersion with a cubic polynomial \citep{watkins2022, ziliotto2025, ziliotto2026},
\begin{equation}
\sigma_{\rm R}(R)=c_0+c_1R+c_2R^2+c_3R^3.
\label{eq:sigmar}
\end{equation}
We require the dispersion to be positive and decreasing with radius, imposing the following conditions:
\begin{equation}
\sigma_{\rm R}(R)>0,\qquad
\frac{{\rm d}\sigma_{\rm R}}{{\rm d}R}=c_1+2c_2R+3c_3R^2<0.
\label{eq:sigma_pos}
\end{equation}
The projected anisotropy is defined as in \citet{dalessandro2024, aros2025, ziliotto2025}.
\begin{equation}
\beta(R)=1-\frac{\sigma_{\rm T}^2(R)}{\sigma_{\rm R}^2(R)}
=\frac{\beta_\infty R^2}{R_{\rm a}^2+R^2}
\left(1-\frac{R}{R_{\rm t}}\right)
\label{eq:ani}
\end{equation}
Here $\beta_\infty$ controls the amplitude, $R_{\rm a}$ sets the radial scale on which anisotropy develops, and $R_{\rm t}$ is the model radius where isotropy is recovered\footnote{Note that the parameter $R_{\rm t}$ is not a measurement of the tidal radius.}. Positive, zero, and negative $\beta$ indicate radial anisotropy, isotropy, and tangential anisotropy, respectively. In practice, the model prescribes the functional forms of $\sigma_{\rm R}(R)$ (Eq.~\ref{eq:sigmar}) and of the anisotropy profile (Eq.~\ref{eq:ani}), while the tangential dispersion is not modelled independently but follows as $\sigma_{\rm T}^2(R)=[1-\beta(R)]\,\sigma_{\rm R}^2(R)$. We require $\beta(R)<1$ over the fitted range and $R_{\rm a}<R_{\rm t}$. The model imposes isotropy at the centre $(R=0)$ and at $R=R_{\rm t}$.

We employed the \textsc{nautilus} Python library \citep{nautilus} to sample the log-likelihood (Eq.~\ref{eq:likelihood}) with respect to $(c_0, c_1, c_2, c_3, \beta_\infty, R_{\rm a}, R_{\rm t})$, using the same broad priors for all populations to avoid introducing any bias. Uncertainties on the best-fit models are then determined from the $16-84^\mathrm{th}$ percentiles of the model posterior distributions. 

\subsection{Energy equipartition}
\label{subsec:equipartition}
For the \jwst\ stellar sample in F3, we study energy equipartition by fitting the two models described in \citet{bianchini2016, watkins2022, aros2023} for the velocity dispersion as a function of stellar mass. The first model assumes a power law
\begin{equation}
\sigma(m)=\sigma_0\left(\frac{m}{m_0}\right)^{-\eta},
\qquad m_0=1\,\msun
\label{eq:powerlaw}
\end{equation}
where $\eta=0$ indicates no mass dependence, while $\eta=0.5$ gives complete equipartition. Values of $\eta<0$ are possible and describe velocity dispersion increasing with stellar mass, or anti-equipartition.

The second formulation was introduced by \citet{bianchini2016} and extended by \citet{aros2023}. Written in terms of the inverse of the equipartition mass, $\mu_\mathrm{eq} = 1/m_\mathrm{eq}$, it reads
\begin{equation}
\sigma(m)=
\begin{cases}
A\, \exp(-0.5\, \mu_\mathrm{eq}\, m), & \mu_\mathrm{eq}\, m \leq 1,\\[3pt]
A\, \exp(-0.5)\, (\mu_\mathrm{eq}\, m)^{-1/2}, & \mu_\mathrm{eq}\, m > 1,
\end{cases}
\label{eq:bianchini}
\end{equation}
where the normalisation $A$ is set so that $\sigma(m_\mathrm{ref}) = \sigma_0$ at the reference mass $m_\mathrm{ref} = 0.5\,\msun$. Stars more massive than $m_\mathrm{eq}$ are in complete equipartition, while less massive stars are only partially equipartitioned. Following \citet{aros2023}, we sample $\mu_\mathrm{eq}$ rather than $m_\mathrm{eq}$, which allows a continuous transition through the case of no mass dependence: $\mu_\mathrm{eq} = 0$ gives a constant dispersion, $\mu_\mathrm{eq} < 0$ gives a dispersion that increases with mass (inverse equipartition), and $\mu_\mathrm{eq} \gtrsim 4\,\msun^{-1}$ (i.e. $m_\mathrm{eq} \lesssim 0.25\,\msun$) corresponds to complete equipartition over the mass range covered by our sample. As for the radial profiles, we fit the radial and tangential components simultaneously and sample the posterior distribution with \textsc{nautilus}. The results of the fitting procedure are presented and discussed in the following.

\section{Results}
\label{sec:results}
In this section we present the results obtained from the fitting procedures described in Sec.~\ref{sec:methods}. We remind the reader that binned results, i.e. coloured points, and best-fit analytical profiles are derived independently, but we present them together as a visual check and comparison. The best-fit parameters and their uncertainties for the radial-profile and equipartition models are listed in Tables~\ref{tab:bestfit_kin} and~\ref{tab:bestfit_eq} in Appendix~\ref{app:bestfit}.

\subsection{Radial kinematic profiles of the multiple populations}
\label{subsec:res_profiles}
\begin{figure*}
    \centering
    \includegraphics[width=0.99\textwidth]{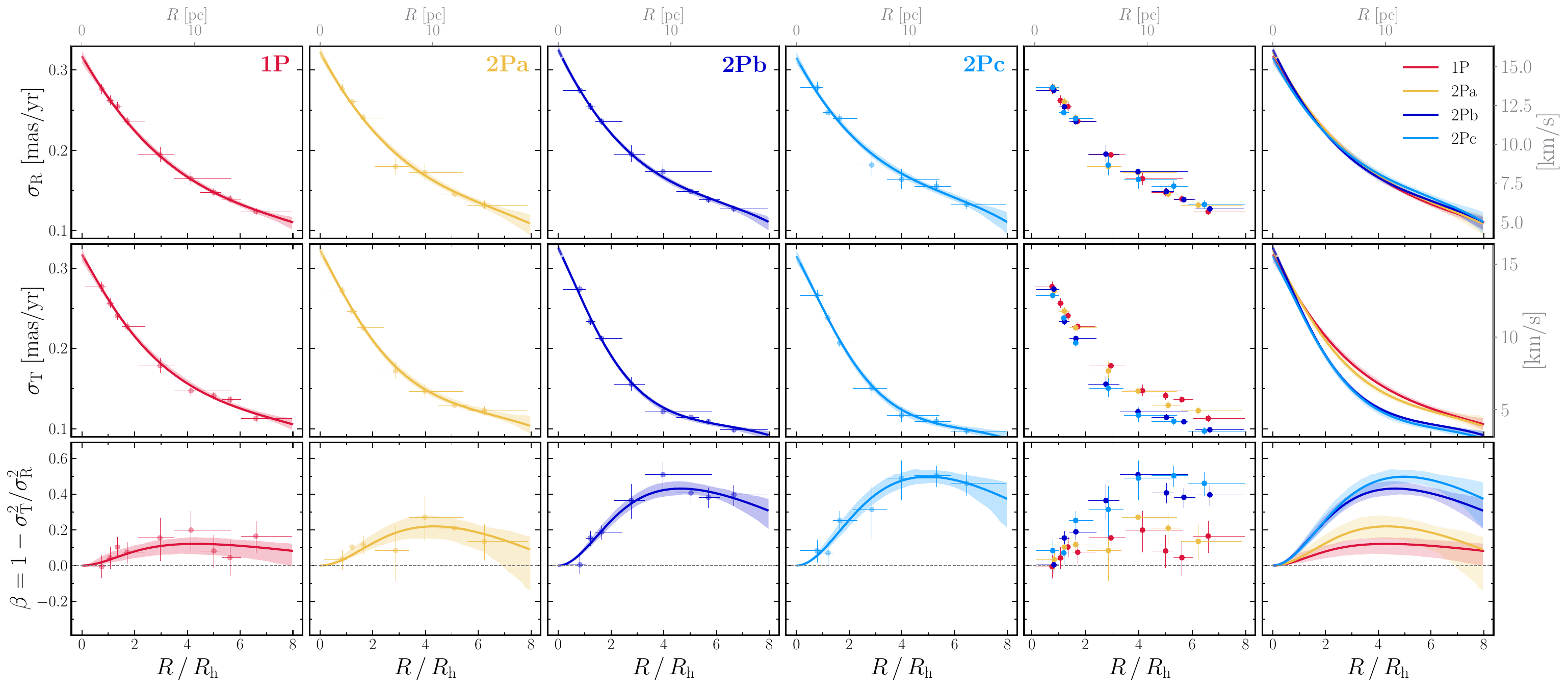}
    \caption{Radial profiles of the four populations. The top, middle, and bottom rows show the radial dispersion, tangential dispersion, and anisotropy. The first four columns show each population separately, and the last two compare the binned estimates and unbinned models. Curves and shaded bands mark posterior medians and $16-84^\mathrm{th}$ percentile intervals. The chemically intermediate 2Pa population displays an anisotropy profile similar to the 1P, while 2Pb and 2Pc are more radially anisotropic. The separation is driven mainly by their lower tangential dispersions.}
    \label{fig:profiles}
\end{figure*}

Figure~\ref{fig:profiles} compares the radial and tangential velocity dispersion profiles and the resulting anisotropy of the four populations, namely 1P, 2Pa, 2Pb and 2Pc, displayed in red, orange, blue and azure, respectively.\footnote{The profiles shown are obtained by fitting both MS and RGB stars. We verified that restricting the fits to MS stars only, which are available in all three fields, gives consistent profiles and the same conclusions. RGB stars alone cannot be used for the same test, as they are only available in F1 and F2.} Visual inspection reveals that all populations share similar sky-projected radial kinematics (top row). However, significant differences occur along the tangential direction (middle row), with 2Pb and 2Pc characterised by a lower tangential velocity dispersion than 1P and 2Pa. As a result, the more chemically enriched populations, 2Pb and 2Pc, exhibit strong radial anisotropy.

On the other hand, the anisotropy profile of the chemically intermediate 2Pa is more similar to that of 1P, with only mild radial anisotropy ($\beta \lesssim 0.2$) across the entire field of view, and remains distinct from those of the more chemically enriched 2Pb and 2Pc. The anisotropy of 2Pa is lower than that of 2Pb and 2Pc by $\gtrsim3\sigma$ at $4$ and $6\,\rh$ (Table~\ref{tab:beta_rh}), while it is consistent with that of 1P within $1.5\sigma$ at all radii. We also note that, in the outermost regions, 2Pc appears more radially anisotropic than 2Pb, although this difference is only marginal and partly depends on the poorly constrained $R_{\rm t}$ (Appendix~\ref{app:bestfit}).

To test whether the 1P-like kinematics of 2Pa could result from misclassified 1P stars, we repeated the fits after replacing an increasing fraction of the 2Pb stars with 1P stars. As shown in Fig.~\ref{fig:contamination}, a contamination above 60 per cent would be needed to reproduce the anisotropy of 2Pa, well above the misclassification expected from the ChMs. The difference between 2Pa and the other 2P groups therefore cannot be explained by contamination in the population selection.

These results are consistent with \citet{bellini2015}, who found similar kinematics for populations A, B and C and lower tangential velocity dispersions for D and E in central \hst\ field. However, the present analysis traces the anisotropy profile of 2Pa separately across the combined field with substantially better definition, while also extending the comparison to M-dwarfs. The same result is recovered for the rMS population, as shown in Appendix~\ref{app:rms} and Fig.~\ref{fig:rms}, with the combined rMS sample also exhibiting only weak radial anisotropy, in contrast to 2Pb and 2Pc.

\subsection{Energy equipartition}
\label{subsec:res_equipartition}
\begin{figure*}
    \centering
    \includegraphics[width=0.99\textwidth]{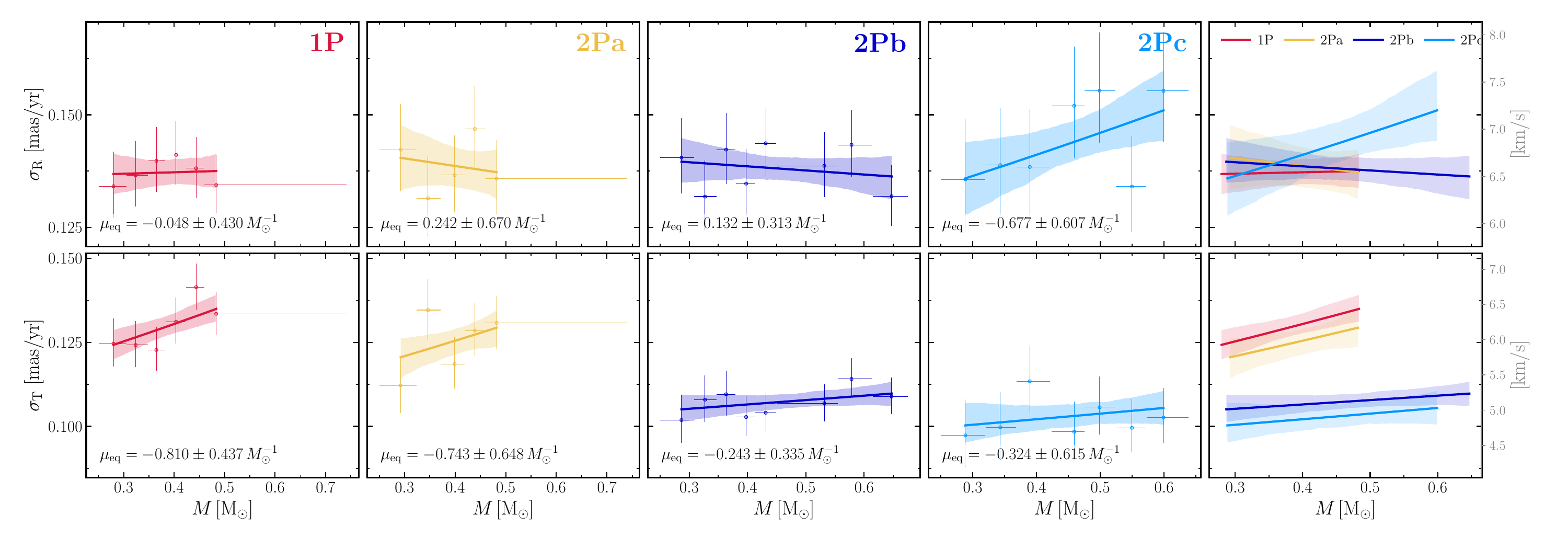}
    \caption{Radial (top) and tangential (bottom) velocity dispersion as a function of stellar mass in the \jwst\ field. The first four columns show the populations separately, and the last is a direct comparison. Points show binned estimates, while curves and shaded bands mark posterior medians and $16-84^\mathrm{th}$ percentile intervals. Each panel gives the corresponding value of $\mu_{\rm eq}$ and uncertainties.}
    \label{fig:equipartition_profiles}
\end{figure*}

\begin{figure*}
    \centering
    \includegraphics[width=0.75\textwidth]{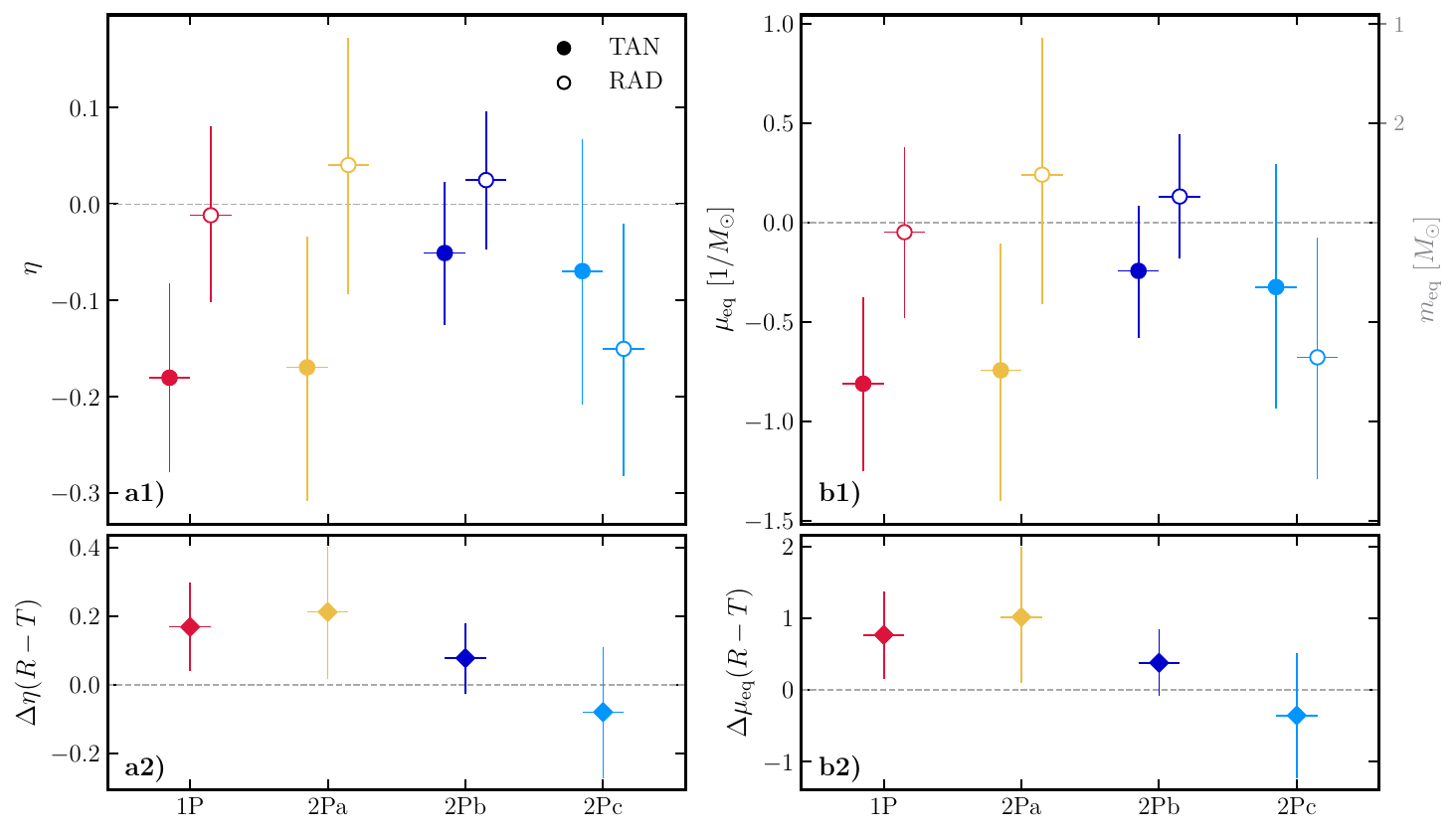}
    \caption{Radial and tangential equipartition parameters of the four populations. Panels a1-b1 show the power-law exponent $\eta$ and the inverse of the equipartition mass $\mu_{\rm eq}$, respectively. Filled and open circles denote tangential and radial motions, respectively, with vertical bars spanning the $16-84^\mathrm{th}$ percentile intervals. Colours match Fig.~\ref{fig:profiles}. The dashed lines indicate no mass dependence. Panels a2-b2 display the difference between the radial and tangential equipartition parameters for each population.}
    \label{fig:equipartition_summary}
\end{figure*}

Figure~\ref{fig:equipartition_profiles} shows the radial and tangential dispersions as functions of stellar mass, together with the best-fit \citet{bianchini2016} models, while Fig.~\ref{fig:equipartition_summary} displays the equipartition parameters of both models (Eqs.~\ref{eq:powerlaw} and \ref{eq:bianchini}), namely the power-law exponent $\eta$ and inverse of the equipartition mass $\mu_\mathrm{eq}$. These measurements describe the mass dependence averaged over the observed F3 field, chosen because of the wider stellar mass range covered by the \jwst\ observations.\footnote{Within F3, the velocity dispersion still decreases by $\sim 30$ per cent with radius (Fig.~\ref{fig:profiles}). We therefore repeated the fits after removing this radial variation, by dividing the radial and tangential proper motions of each star and their uncertainties by the dispersion profile of its population at the star's radius. The resulting $\eta$ and $\mu_{\rm eq}$ differ by less than $0.5\sigma$ from those presented here, and all the results discussed in this section and in Sec.~\ref{subsec:disc_equipartition} remain unchanged.}

The mass dependence is weak for all four populations. In the radial component, $\eta$ and $\mu_{\rm eq}$ are consistent with zero, with only 2Pc displaying a marginally negative value, while in the tangential component the values are negative for all populations, although each only at the $\sim1$--$2\sigma$ level. This suggests a weak inverse mass dependence, with the tangential dispersion slightly increasing with stellar mass. The difference between the radial and tangential parameters (Fig.~\ref{fig:equipartition_summary}, panels a2 and b2) is similar for 1P and 2Pa, decreases towards 2Pb, and becomes negative for 2Pc. These trends are consistent across both models, although the large uncertainties prevent a firm detection.
Since the mass dependence is sensitive to the proper-motion uncertainties of the faintest stars, we repeated the fits after rescaling all uncertainties by factors between 0.5 and 1.5: although the individual parameters shift by up to $\sim1\sigma$ for rescalings of $\pm20$ per cent, the relative behaviour of the populations is unchanged. In all cases, the radial$-$tangential difference remains positive for 1P and 2Pa and decreases towards 2Pb and 2Pc, and the inverse mass dependence persists in the tangential component of 1P and 2Pa and in the radial component of 2Pc.

\section{Discussion}
\label{sec:discussion and conclusions}

\subsection{Interpreting the kinematics of 2Pa}
\label{subsec:disc_profile}

The radial profiles in Fig.~\ref{fig:profiles} show that 2Pb and 2Pc are radially anisotropic, while 2Pa exhibits only mild radial anisotropy, similar to that of the 1P. In simulations where the 2P forms in a compact subsystem embedded in a more extended 1P, the 2P evolves like a tidally underfilling system and develops radial anisotropy, mainly through a lower tangential dispersion, whereas the 1P behaves like a tidally filling system and remains close to isotropy \citep{tiongco2019, vesperini2021}. The kinematics of 2Pb and 2Pc are therefore consistent with a compact initial configuration, as found for 2P stars in other clusters \citep{dalessandro2024, cordoni2025a}. In contrast, the 1P-like kinematics of 2Pa suggest that it started from different initial conditions from those of 2Pb and 2Pc, possibly with a different concentration or initial kinematics.

\begin{figure*}
    \centering
    \includegraphics[width=0.99\textwidth, trim={0 0 2cm 0}, clip]{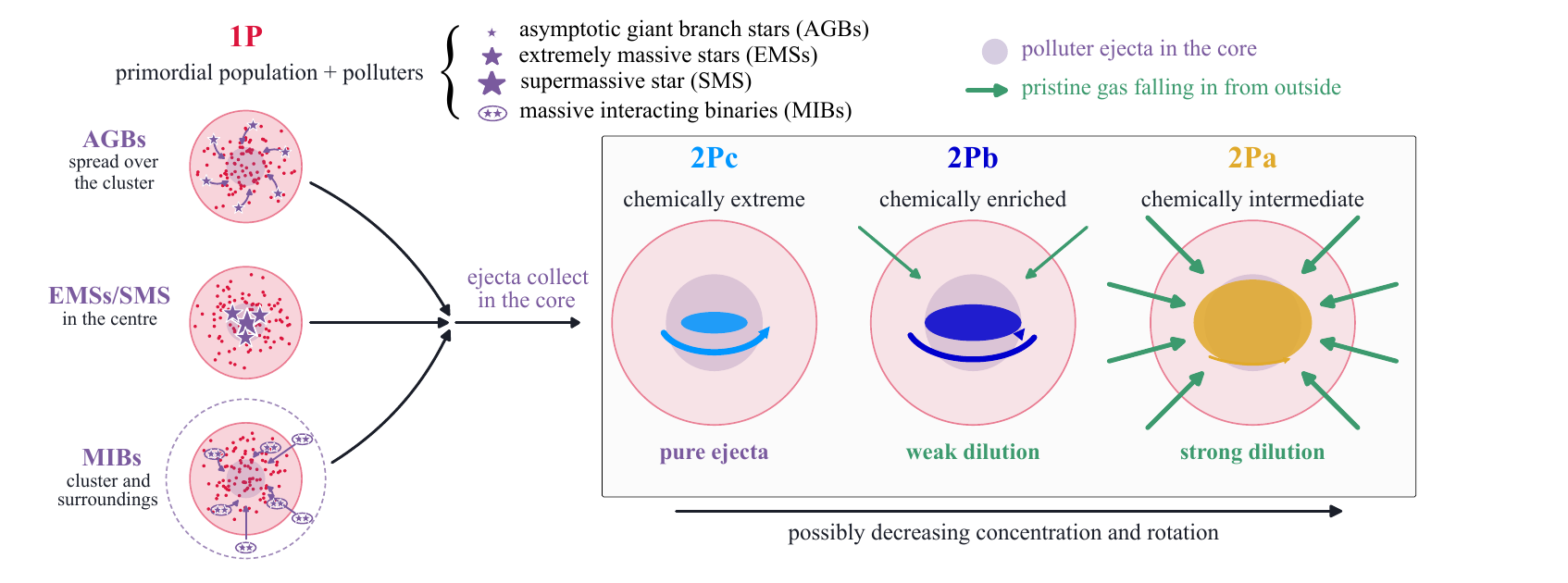}
    \caption{Schematic view of the dilution picture discussed in Sec.~\ref{subsec:disc_profile}. Left: the 1P with its polluters, either distributed throughout the cluster (AGB stars), concentrated in the centre (a single SMS or several EMSs), or in the cluster and the star-forming region around it (MIBs). We represent SMS and EMSs together for visualisation purposes only. In all cases, the ejecta collect in the cluster core. Right: the 2P groups form from ejecta increasingly diluted with pristine gas falling in from outside, from pure ejecta (2Pc) to strong dilution (2Pa). This could result in progressively less centrally concentrated and more slowly rotating 2P groups, as found in hydrodynamical simulations \citep{calura2019, lacchin2022}. The groups are ordered by the degree of dilution, not by formation time, which depends on the scenario. The sketch illustrates one possible origin of the different initial configurations discussed in the text; it is qualitative and not to scale.}
    \label{fig:hyp_dilution}
\end{figure*}

Such a different initial configuration could be related to the degree of dilution of the gas from which the 2P groups formed. The chemical composition of 2Pa is closer to that of the 1P than those of 2Pb and 2Pc, suggesting a larger contribution of pristine gas to its formation \citep{dercole2008, dantona2016}. We illustrate this picture qualitatively in Fig.~\ref{fig:hyp_dilution}, with the aim of discussing how the kinematics of the 2P groups could fit within the different scenarios, rather than identifying the polluters or favouring any of them. Whether the polluters are distributed throughout the cluster and possibly in its surroundings, as AGB stars or MIBs, or concentrated in its centre, as a single SMS or several EMSs, their ejecta are expected to collect in the cluster core \citep{dercole2008, demink2009, gieles2018, renzini2022, gieles2025}. 2P groups forming from gas with different degrees of dilution could therefore start from different initial conditions. In hydrodynamical simulations of 2P formation from AGB ejecta diluted with infalling pristine gas, the most helium-rich stars form in the most compact configuration, while the less enriched 2P stars are more extended \citep{calura2019, lacchin2022} and, in some models, also form with lower rotation \citep{lacchin2022}. A dependence of the initial conditions on dilution could also be expected in the scenario of \citet{renzini2022}, in which the cluster forms inside a dwarf galaxy that also supplies most of the processed material, and MIB ejecta and residual gas keep flowing into the centre of the cluster, forming the 2P in successive bursts. A similar dependence could be expected in SMS and EMS scenarios, where the most extreme compositions could be expected closer to the cluster centre \citep{gieles2018, gieles2025}.\footnote{To our knowledge, no hydrodynamical simulations following the formation and spatial distribution of 2P stars in the SMS and EMS scenarios are currently available.}

Such possibilities are displayed in Fig.~\ref{fig:hyp_dilution}, where the sequence from 2Pc to 2Pa mainly reflects the degree of dilution and the possible resulting initial configuration, not a formation chronology. In AGB and MIB scenarios, the first 2P stars form from almost pure ejecta \citep{dercole2008, dantona2016, renzini2022}, whereas in SMS and EMS scenarios the enrichment of the ejecta increases with time \citep{gieles2018, gieles2025}. In all these cases, regardless of the formation order, 2P groups formed from gas with different degrees of dilution could start from different initial conditions and therefore evolve towards different present-day kinematics.

Besides the interplay between polluters' ejecta and pristine gas, dynamical simulations have shown that the tidal field plays an important role in shaping the spatial distribution \citep{vesperini2013} and the present-day kinematics of cluster stars, and of 1P and 2P in particular \citep{tiongco2016a, tiongco2019, vesperini2021}. If the 2P groups formed over a timescale comparable to the variations of the tidal field, for instance along an eccentric orbit or in a rapidly changing tidal field, they may have formed under different tidal conditions, which could have affected their initial configuration. Since a cluster adapts to a new tidal environment within a few relaxation times \citep{bianchini2017}, such an effect would survive only through the structure of each group at birth. This possibility cannot be tested with the present-day orbit of NGC\,2808 \citep{baumgardt2019, vasiliev2021}, as the cluster likely formed within the Gaia-Enceladus progenitor \citep{massari2019}, where its early tidal history was set, and would require dedicated simulations in a time-dependent tidal field \citep[see e.g.][]{dellacroce2026}.

Overall, similar analyses of other clusters hosting chemically intermediate 2P groups are  needed to establish how general the behaviour of NGC\,2808 is. The picture is likely more complex than outlined above, as in 47\,Tuc the chemically intermediate 2P stars (2G$_{\rm A}$ in their notation) share the kinematics of the most enriched ones (2G$_{\rm B}$) rather than those of the 1P \citep{ziliotto2025}. A larger sample of clusters with different masses, orbits, and dynamical states will be essential to understand the origin of these differences.

\subsection{Energy equipartition in the outer field F3}
\label{subsec:disc_equipartition}
The weak degree of equipartition is consistent with the location of the \jwst\ field in the outer regions of the cluster, where the relaxation times are long \citep{livernois2024}. The tangential equipartition parameters are negative for all populations, and those of 1P and 2Pa, together with the radial ones of 2Pc, remain negative for all the rescalings of the proper-motion uncertainties discussed in Sec.~\ref{subsec:res_equipartition}. This may represent tentative evidence of inverted equipartition, in which lower-mass stars have smaller velocity dispersions than higher-mass ones, as found by \citet{pavlik2022} in the outer regions of initially isotropic systems. 

The relative degree of equipartition of the radial and tangential components depends on the initial kinematics of the system. In the intermediate and outer regions, initially isotropic systems evolve faster towards equipartition in the radial component, while initially radially anisotropic systems evolve faster in the tangential one \citep{pavlik2022}, as do rotating systems \citep{livernois2022}. In our data, the difference between the radial and tangential equipartition parameters tentatively decreases from 1P to 2Pc. In the multiple-population models of \citet{livernois2024}, equipartition generally proceeds faster in the tangential component, particularly for the second generation, as we find for 2Pc. Moreover, in their outer regions (see e.g. their Fig.~12), the difference between the radial and tangential equipartition parameters is positive for the first generation and negative or null for the second, the same sign pattern we find between 1P and 2Pc. However, given the weak mass dependence and the uncertainties, we refrain from drawing firm conclusions. 

\section{Summary and conclusions}
\label{sec:summary}
In this work, we combined \jwst\ and \hst\ photometry and proper motions to investigate the internal dynamics of multiple populations in NGC\,2808, here labelled 1P, 2Pa, 2Pb, and 2Pc. We measured their velocity dispersion and anisotropy profiles across the combined field, extending the comparison from RGB to M-dwarf stars, and investigated energy equipartition in the outer \jwst\ field.

The chemically intermediate 2Pa population exhibits an anisotropy profile similar to that of the 1P, with only mild radial anisotropy, whereas 2Pb and 2Pc are strongly radially anisotropic. This separation is driven by the lower tangential dispersions of 2Pb and 2Pc, while the radial dispersions are similar for all four populations. This is the first characterisation of the kinematics of 2Pa separately from the 1P out to about eight half-light radii and down to M-dwarfs.

The equipartition analysis reveals only a weak mass dependence of the velocity dispersion, with tentative evidence of inverted equipartition. The difference between the radial and tangential equipartition parameters is similar for 1P and 2Pa and decreases towards 2Pc, although this trend remains tentative given the uncertainties. Since models predict that the relative degree of equipartition of the two components depends on the initial kinematics of each population \citep{pavlik2022, livernois2022, livernois2024}, measurements of this quantity can provide additional constraints on the formation of multiple populations.

Our main finding is that the 2P of NGC\,2808 is not a dynamically homogeneous component, and chemically distinct 2P groups have different kinematics. The 1P-like kinematics of 2Pa suggest that the 2P groups did not all form under the same conditions or did not experience the same dynamical evolution. In particular, 2Pa may have started from initial conditions different from those of 2Pb and 2Pc. Formation scenarios must therefore reproduce not only the chemical composition of the different 2P groups, but also their different kinematics. Dedicated simulations following the formation and long-term dynamical evolution of the individual 2P groups, together with similar analyses of other clusters, are now needed to identify the origin of these differences.

\section*{Acknowledgements}
GC acknowledges support from the Australian Research Council (ARC) through the Discovery Early Career Researcher Award DE260100717. PB acknowledges financial support from the IdEx framework of the University of Strasbourg. YC acknowledges support from the grant RYC2021-032718-I, financed by MCIN/AEI/10.13039/501100011033 and the European Union NextGenerationEU/PRTR. ALV acknowledges support from a UKRI Future Leaders Fellowship (MR/S018859/1; MR/X011097/1). This work is based on observations made with the NASA/ESA/CSA \textit{James Webb Space Telescope}, associated with program GO-8960, and with the NASA/ESA \textit{Hubble Space Telescope}. The data were obtained from the Mikulski Archive for Space Telescopes at the Space Telescope Science Institute, which is operated by the Association of Universities for Research in Astronomy, Inc., under NASA contracts NAS 5-03127 for \jwst\ and NAS 5-26555 for \hst. This work made use of \textsc{nautilus} \citep{nautilus}, \textsc{dynesty} \citep{dynesty}, and \textsc{scikit-learn} \citep{scikit-learn}.

\section*{Data Availability}
The data underlying this article will be shared on reasonable request to the corresponding author.

\bibliographystyle{mnras}

\bibliography{main}

\appendix

\section{Selection of multiple populations}
\label{app:chm_selection}
In this appendix, we describe how the populations were selected on the \jwst\ ChM of F3 and the tests we performed to assess the robustness of this selection. We show only the \jwst\ selection, as it is presented here for the first time, while the \hst\ selections follow the published procedure for NGC\,2808 \citep{milone2015}. We performed three tests: we removed stars close to the population boundaries (Sec.~\ref{app:chm_borders}), we replaced our selection with a Gaussian mixture model (Sec.~\ref{app:chm_gmm}), and we estimated the 1P contamination needed to make 2Pb appear as 2Pa (Sec.~\ref{app:contamination}). The results of these tests support the conclusions of Sec.~\ref{sec:discussion and conclusions}. Finally, Sec.~\ref{app:upperms} describes the selection of the populations along the bright MS.

\subsection{Population selection on the \jwst\ chromosome map}
\label{app:chm_kde}
\begin{figure*}
    \centering
    \includegraphics[width=0.7\textwidth]{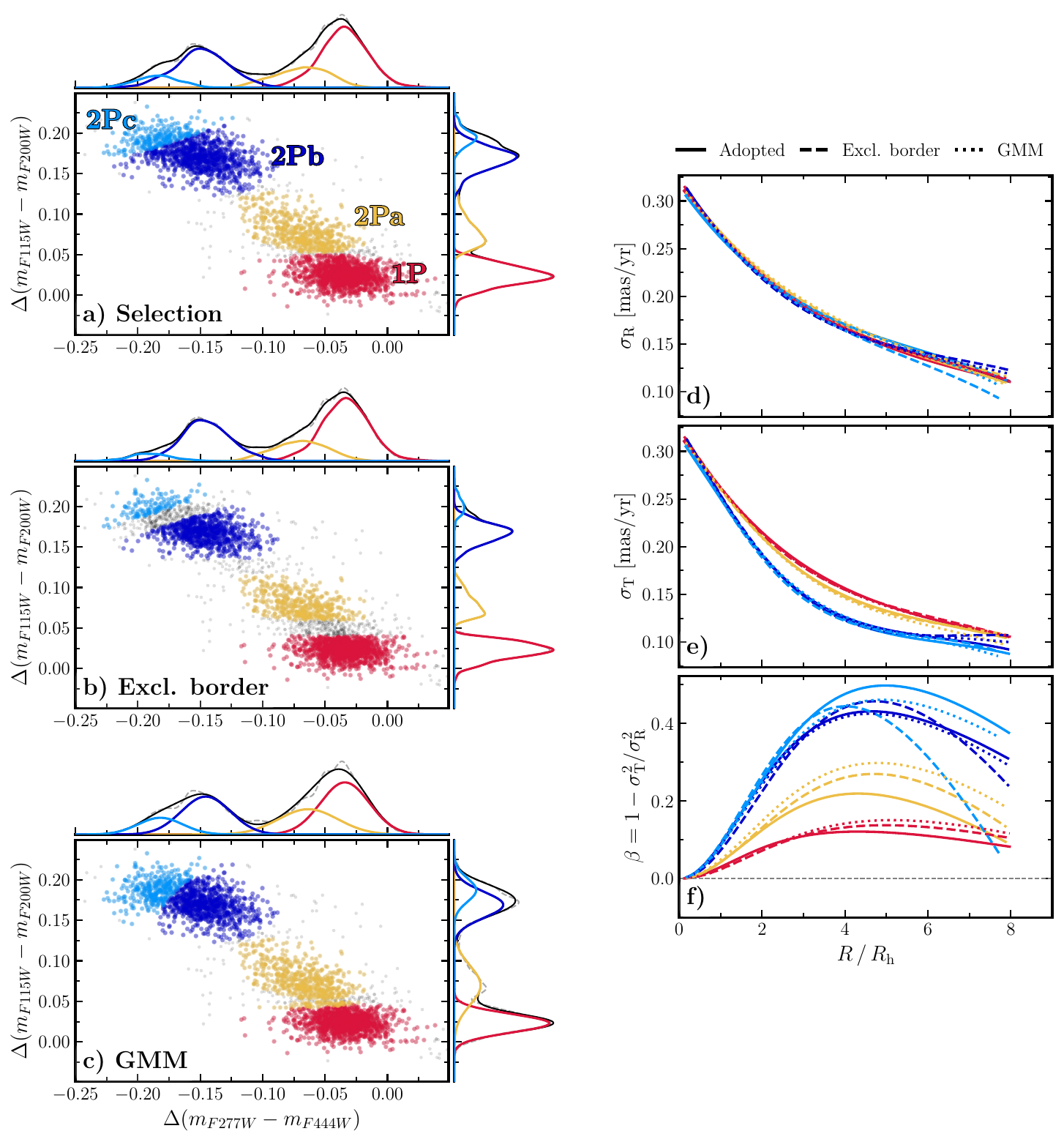}
    \caption{Population selection on the \jwst\ M-dwarf ChM of F3. Colours identify 1P, 2Pa, 2Pb, and 2Pc, while excluded stars are grey. The marginal panels show the $x$ and $y$ KDEs of the full sample (black) and of the selected populations (coloured). Panel a) shows the selection used throughout this work, panel b) the selection after removing stars close to the boundaries, and panel c) the GMM selection. Panels d), e), and f) show the resulting $\sigma_{\rm R}$, $\sigma_{\rm T}$, and $\beta$ radial profiles of each population, obtained with the selections of panels a) (solid lines), b) (dashed lines), and c) (dotted lines), applied to both the \hst\ and \jwst\ samples. Uncertainties are omitted for readability, and are consistent with those in Fig.~\ref{fig:profiles}.}
    \label{fig:chm_selection}
\end{figure*}

Figure~\ref{fig:chm_selection}a illustrates the selection, showing the M-dwarf ChM and the kernel density estimates (KDEs) in the marginal panels. Black curves refer to all stars, while coloured curves show the selected populations. We identified 1P, 2Pa, 2Pb, and 2Pc in order of increasing $\Delta(\filt{115},\filt{200})$, corresponding to increasing light-element processing. The 1P/2Pa and 2Pa/2Pb boundaries are horizontal lines at $\Delta(\filt{115},\filt{200})=0.05$ and $0.125$\,mag. The 2Pb and 2Pc groups partly overlap in this coordinate but separate in $\Delta(\filt{277},\filt{444})$, so we divide them with an inclined line (see also the 2D histogram in Fig.~\ref{fig:photometry jwst}e, where 2Pb and 2Pc are clearly visible).

\subsection{Removing stars close to the population boundaries}
\label{app:chm_borders}
Stars close to the population boundaries are the most likely to be misclassified. We therefore repeated the analysis after excluding stars within $\pm0.01$\,mag of the boundaries on the \jwst\ ChM, and within $\pm0.03$\,mag on the \hst\ ChMs of RGB and MS stars. The distance from each boundary is measured perpendicular to the dividing line.

We repeated the complete dynamical analysis, including binned estimates, radial-profile fits, and equipartition fits. The results agree within the uncertainties with those of the adopted selection, and the profiles (dashed lines in Fig.~\ref{fig:chm_selection}d--f) show only small differences: 2Pa remains consistent with 1P, while 2Pb and 2Pc remain more radially anisotropic. The only noticeable change concerns 2Pc, whose outermost profile is less constrained, as only a few 2Pc stars remain after the cut.

\subsection{Comparison with Gaussian mixture models}
\label{app:chm_gmm}
We also fitted a four-component GMM with full covariance matrices to the $(\Delta(\filt{277},\filt{444}),\Delta(\filt{115},\filt{200}))$ pairs, using \textsc{GaussianMixture} \citep{scikit-learn}, and assigned each star to the component with the highest probability. Stars with large values of the two ChM coordinates were removed before the fit, which is why some stars in Fig.~\ref{fig:chm_selection}c are not assigned to any population. The same procedure was applied to the \hst\ ChMs of F1 and F2, separately for RGB and MS stars. The mixture components follow the KDE-defined regions, and the two classifications agree for most stars, differing mainly near the boundaries, where adjacent components have comparable probabilities. The resulting profiles (dotted lines in Fig.~\ref{fig:chm_selection}d--f) are consistent with those of the adopted selection, and the separation between the 1P/2Pa and 2Pb/2Pc anisotropy profiles is preserved.

\subsection{Contamination by 1P stars}
\label{app:contamination}
\begin{figure}
    \centering
    \includegraphics[width=0.45\textwidth]{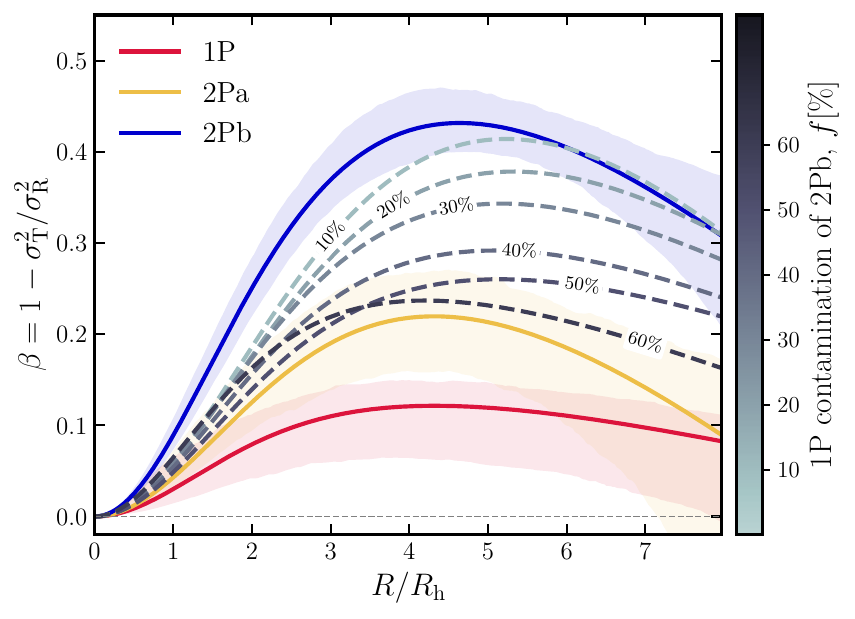}
    \caption{Anisotropy profiles of 1P, 2Pa, and 2Pb (solid lines), compared with those of the 2Pb sample in which a fraction $f$ of the stars has been replaced with 1P stars (dashed lines, colour-coded by $f$). Only for $f\gtrsim60$ per cent does the contaminated profile become consistent with that of 2Pa.}
    \label{fig:contamination}
\end{figure}

Since 1P and 2Pa are adjacent on the ChMs, misclassified 1P stars could in principle make 2Pa appear more 1P-like. To test whether this could explain the kinematics of 2Pa, we estimated how much 1P contamination would be needed to make an intrinsically 2Pb-like population appear as 2Pa. In each field, we replaced a fraction $f$ of the 2Pb stars with randomly drawn 1P stars, keeping the number of stars fixed, and refitted the radial profiles as described in Sec.~\ref{sec:methods}. As shown in Fig.~\ref{fig:contamination}, the anisotropy of the contaminated sample decreases only gradually with $f$, and becomes consistent with that of 2Pa only for $f\gtrsim60$ per cent. Such a contamination is far larger than any plausible misclassification on the ChMs, so contamination by 1P stars cannot explain the anisotropy profile of 2Pa.

\subsection{Populations along the bright main sequence}
\label{app:upperms}
\begin{figure*}
    \centering
    \includegraphics[width=0.90\linewidth]{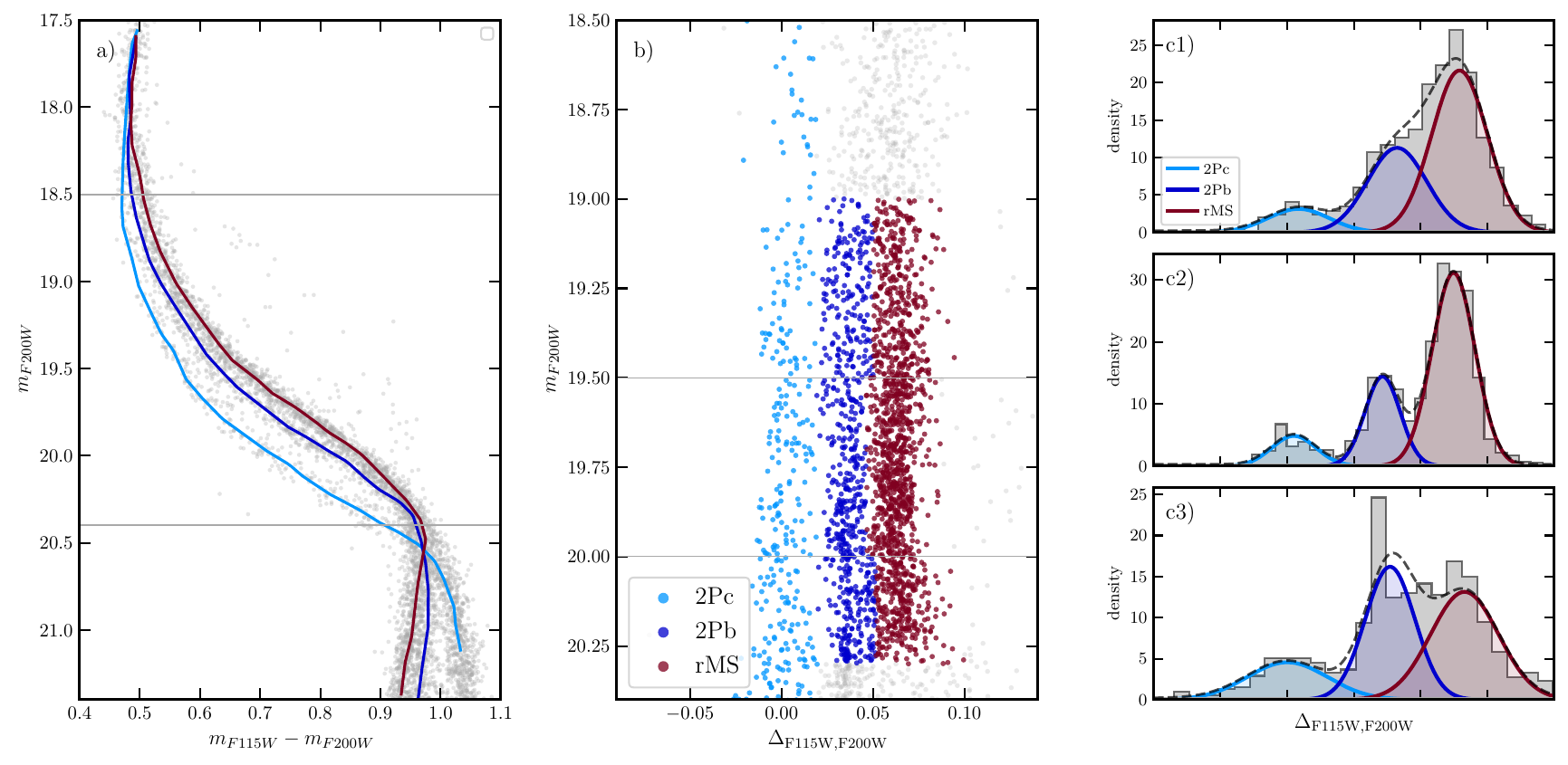}
    \caption{Bright-MS selection of rMS, 2Pb, and 2Pc stars, shown in burgundy, blue, and azure, respectively. Panel a) shows the CMD of Fig.~\ref{fig:photometry jwst}d with the three fiducials used for the verticalisation. In panel b), the fiducials map onto $0$, $w_{\rm L}$, and $w_{\rm L}+w_{\rm R}$, with widths measured at $\filt{200}=19.0$. Panels c1--c3) show the Gaussian-mixture fits in the three magnitude intervals, from bright to faint. Grey horizontal lines mark the magnitude range used for the verticalisation (panel a) and the three magnitude intervals of the GMM fits (panel b).}
    \label{fig:upperms_gmm}
\end{figure*}

Above the MS knee, the \jwst\ colours resolve the rMS, which combines 1P and 2Pa, and the helium-rich 2Pb and 2Pc sequences. We selected these three groups in the \cmd{200, 115, 200} CMD of Fig.~\ref{fig:photometry jwst}d, following the procedure shown in Fig.~\ref{fig:upperms_gmm}. We first traced the blue edge, intermediate sequence, and red edge of the bright MS with three fiducial lines, and verticalised the CMD so that the fiducials correspond to $0$, $w_{\rm L}$, and $w_{\rm L}+w_{\rm R}$, where $w_{\rm L}$ and $w_{\rm R}$ are the colour separations between adjacent fiducials at $\filt{200}=19.0$. This preserves the relative spacing of the three sequences. We then fitted the verticalised colour distributions in the intervals $18.5\leq\filt{200}<19.5$, $19.5\leq\filt{200}<20.0$, and $20.0\leq\filt{200}\leq20.5$ with three Gaussian components using \textsc{GaussianMixture} \citep{scikit-learn}. The three components correspond to 2Pc, 2Pb, and rMS in order of increasing verticalised colour, with means and dispersions consistent across the magnitude intervals, and each star was assigned to the component with the highest membership probability. The rMS sample defined here is the one used in Appendix~\ref{app:rms}, while the bright-MS 2Pb and 2Pc stars are combined with the corresponding M-dwarfs selected on the \jwst\ ChM.

\section{Dynamical properties of the red main sequence}
\label{app:rms}
Earlier studies grouped the helium-normal upper-MS stars into the red main sequence \citep[rMS;][]{piotto2007}. Our \hst\ and \jwst\ photometry separates this sequence into the 1P and 2Pa populations, but we retain the combined rMS sample here to facilitate comparison with the historical classification.

In F3, the rMS sample is not simply the union of 1P and 2Pa. Above the MS knee, the rMS can be followed along the whole bright MS over the entire field (Appendix~\ref{app:upperms}), whereas 1P and 2Pa can be separated only in the small magnitude interval and area covered by the \hst\ F2 observations. The rMS therefore include more stars and cover a wider mass range, which explains its smaller uncertainties in Table~\ref{tab:bestfit_eq}.

\begin{figure*}
    \centering
    \includegraphics[width=0.70\linewidth]{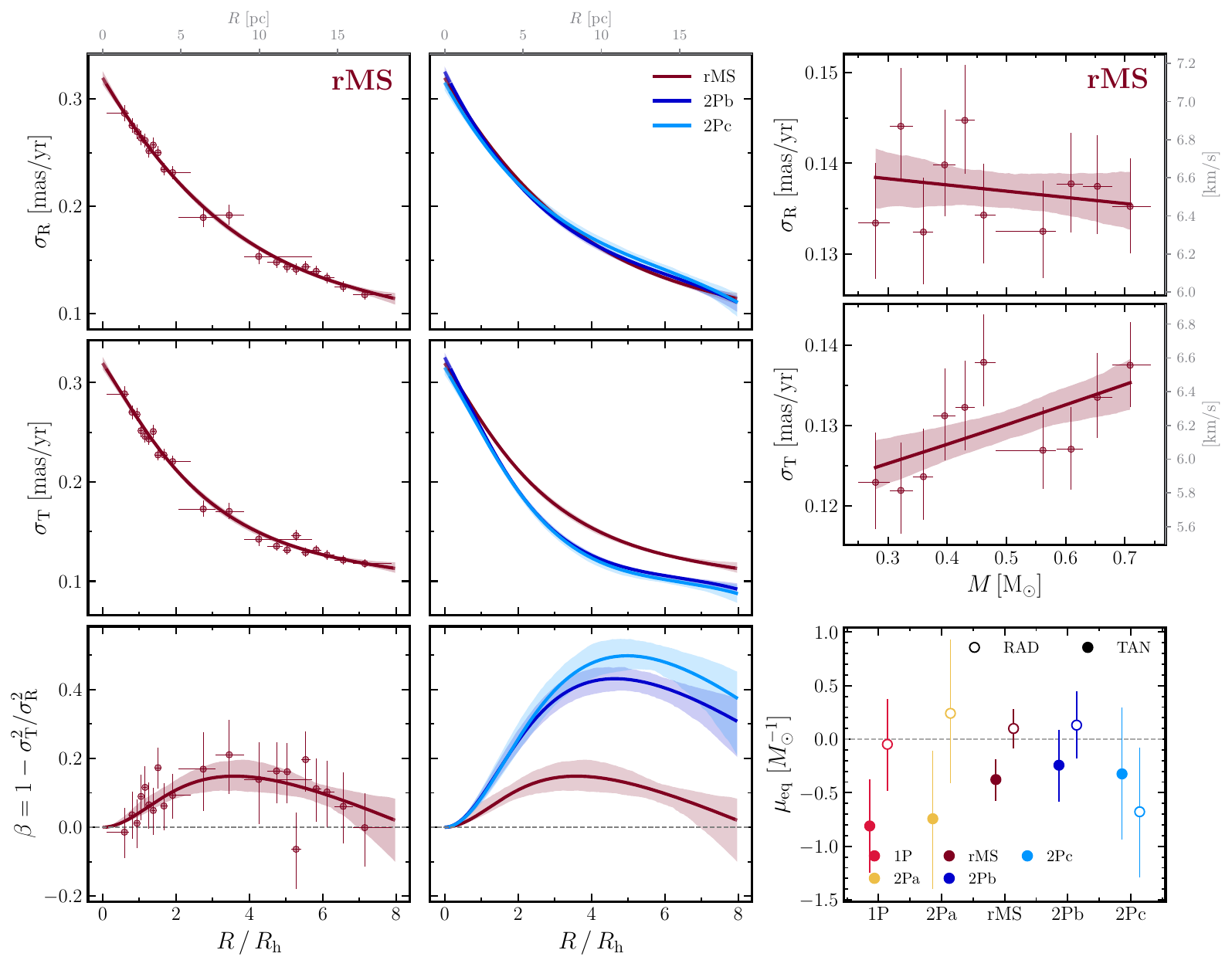}
    \caption{Internal dynamics of the rMS compared with the four populations analysed in this work. The left column shows the binned radial and tangential velocity dispersions and anisotropy, together with the unbinned models. The middle column compares the rMS, 2Pb, and 2Pc best-fit profiles. The upper and middle panels of the right column show the rMS dispersions as functions of stellar mass, while the lower-right panel compares the radial and tangential $\mu_{\rm eq}$ of 1P, 2Pa, rMS, 2Pb, and 2Pc. The rMS follows 1P and 2Pa, as expected, since it combines these two groups.}
    \label{fig:rms}
\end{figure*}

The radial-dispersion and anisotropy profiles of the rMS follow those of 1P and 2Pa, and its equipartition parameters are consistent with those of 1P and 2Pa. The combined sample therefore recovers the earlier classification, while separating 1P and 2Pa reveals the behaviour of the chemically intermediate population.

\section{Best-fit parameters}
\label{app:bestfit}
To allow the profiles to be reproduced, Tables~\ref{tab:bestfit_kin} and~\ref{tab:bestfit_eq} list the posterior medians and $16-84^\mathrm{th}$ percentile intervals of the parameters of the radial-profile (Eqs.~\ref{eq:sigmar} and~\ref{eq:ani}) and equipartition (Eqs.~\ref{eq:powerlaw} and~\ref{eq:bianchini}) models, for the four populations and the rMS.

The data do not constrain $R_{\rm t}$: the posteriors are broad, with medians close to 10\,arcmin for 1P, 2Pb, and 2Pc, well beyond the outermost data point at $R\simeq6.4$\,arcmin ($8\,\rh$). The profiles beyond this radius are therefore extrapolations, and the anisotropy close to the edge of the field partly depends on $R_{\rm t}$. Table~\ref{tab:beta_rh} lists $\beta$ at fixed radii, derived directly from the posterior draws.

\begin{table*}
\setlength{\tabcolsep}{4pt}
\centering
\begin{tabular}{lccccccc}
\hline
Population & $c_0$ & $c_1$ & $c_2$ & $c_3$ & $\beta_\infty$ & $R_{\rm a}$ & $R_{\rm t}$ \\
 % & [\masyr] & [$10^{-2}$\,\masyr\,arcmin$^{-1}$] & [$10^{-3}$\,\masyr\,arcmin$^{-2}$] & [$10^{-4}$\,\masyr\,arcmin$^{-3}$] & & [arcmin] & [arcmin] \\
\hline
1P & $0.3173^{+0.0083}_{-0.0082}$ & $-7.19^{+1.11}_{-1.09}$ & $9.35^{+3.63}_{-3.54}$ & $-4.96^{+3.34}_{-3.60}$ & $0.25^{+0.37}_{-0.15}$ & $2.08^{+2.30}_{-1.41}$ & $9.98^{+5.92}_{-3.82}$ \\
2Pa & $0.3225^{+0.0076}_{-0.0084}$ & $-7.76^{+1.20}_{-1.01}$ & $11.47^{+3.48}_{-3.94}$ & $-7.15^{+3.84}_{-3.64}$ & $0.70^{+1.03}_{-0.41}$ & $2.95^{+1.82}_{-1.49}$ & $7.55^{+6.61}_{-1.85}$ \\
2Pb & $0.3253^{+0.0054}_{-0.0062}$ & $-8.55^{+0.88}_{-0.61}$ & $14.23^{+2.01}_{-3.04}$ & $-9.56^{+3.07}_{-2.09}$ & $0.98^{+0.57}_{-0.27}$ & $2.42^{+0.81}_{-0.50}$ & $9.95^{+4.97}_{-2.37}$ \\
2Pc & $0.3152^{+0.0072}_{-0.0081}$ & $-7.67^{+1.18}_{-0.89}$ & $12.75^{+2.99}_{-3.97}$ & $-9.00^{+4.05}_{-3.12}$ & $1.24^{+1.03}_{-0.39}$ & $2.81^{+1.26}_{-0.70}$ & $9.97^{+5.15}_{-2.63}$ \\
rMS & $0.3196^{+0.0066}_{-0.0064}$ & $-7.30^{+0.85}_{-0.90}$ & $9.35^{+2.87}_{-2.66}$ & $-4.64^{+2.50}_{-2.70}$ & $0.41^{+0.63}_{-0.25}$ & $2.20^{+1.79}_{-1.26}$ & $6.75^{+6.31}_{-1.26}$ \\
\hline
\end{tabular}
\caption{Posterior medians and $16-84^\mathrm{th}$ percentile intervals of the radial-profile parameters (Eqs.~\ref{eq:sigmar} and~\ref{eq:ani}), fitted to the combined \hst\ and \jwst\ sample.}
\label{tab:bestfit_kin}
\end{table*}

\begin{table}
\centering
\begin{tabular}{lcccc}
\hline
Population & $1\,\rh$ & $2\,\rh$ & $4\,\rh$ & $6\,\rh$ \\
\hline
1P  & $0.03^{+0.03}_{-0.02}$ & $0.08^{+0.04}_{-0.03}$ & $0.11^{+0.05}_{-0.05}$ & $0.09^{+0.05}_{-0.05}$ \\
2Pa & $0.05^{+0.02}_{-0.02}$ & $0.13^{+0.04}_{-0.04}$ & $0.21^{+0.07}_{-0.06}$ & $0.17^{+0.06}_{-0.07}$ \\
2Pb & $0.09^{+0.02}_{-0.02}$ & $0.25^{+0.03}_{-0.03}$ & $0.43^{+0.04}_{-0.04}$ & $0.41^{+0.03}_{-0.04}$ \\
2Pc & $0.09^{+0.02}_{-0.02}$ & $0.26^{+0.04}_{-0.04}$ & $0.48^{+0.05}_{-0.05}$ & $0.48^{+0.04}_{-0.05}$ \\
rMS & $0.05^{+0.02}_{-0.02}$ & $0.11^{+0.03}_{-0.03}$ & $0.14^{+0.05}_{-0.05}$ & $0.09^{+0.03}_{-0.03}$ \\
\hline
\end{tabular}
\caption{Projected anisotropy $\beta$ at fixed radii: posterior medians and $16-84^\mathrm{th}$ percentile intervals, computed from the posterior draws of the radial-profile fits.}
\label{tab:beta_rh}
\end{table}

\begin{table*}
\centering
\begin{tabular}{lcccccccc}
\hline
 & \multicolumn{4}{c}{Power law (Eq.~\ref{eq:powerlaw})} & \multicolumn{4}{c}{\citet{bianchini2016} (Eq.~\ref{eq:bianchini})} \\
Population & $\sigma_{0,{\rm R}}$ & $\eta_{\rm R}$ & $\sigma_{0,{\rm T}}$ & $\eta_{\rm T}$ & $\sigma_{0,{\rm R}}$ & $\mu_{\rm eq,R}$ & $\sigma_{0,{\rm T}}$ & $\mu_{\rm eq,T}$ \\
 % & [\masyr] & & [\masyr] & & [\masyr] & [$\msun^{-1}$] & [\masyr] & [$\msun^{-1}$] \\
\hline
1P & $0.1387^{+0.0127}_{-0.0116}$ & $-0.01^{+0.09}_{-0.09}$ & $0.1545^{+0.0155}_{-0.0139}$ & $-0.18^{+0.10}_{-0.10}$ & $0.1376^{+0.0043}_{-0.0039}$ & $-0.05^{+0.43}_{-0.43}$ & $0.1359^{+0.0042}_{-0.0039}$ & $-0.81^{+0.44}_{-0.44}$ \\
2Pa & $0.1334^{+0.0180}_{-0.0154}$ & $0.04^{+0.13}_{-0.13}$ & $0.1471^{+0.0205}_{-0.0178}$ & $-0.17^{+0.14}_{-0.14}$ & $0.1370^{+0.0058}_{-0.0058}$ & $0.24^{+0.69}_{-0.65}$ & $0.1302^{+0.0054}_{-0.0051}$ & $-0.74^{+0.64}_{-0.66}$ \\
2Pb & $0.1353^{+0.0087}_{-0.0081}$ & $0.03^{+0.07}_{-0.07}$ & $0.1119^{+0.0071}_{-0.0067}$ & $-0.05^{+0.07}_{-0.07}$ & $0.1377^{+0.0028}_{-0.0027}$ & $0.13^{+0.31}_{-0.31}$ & $0.1078^{+0.0023}_{-0.0022}$ & $-0.24^{+0.33}_{-0.34}$ \\
2Pc & $0.1623^{+0.0197}_{-0.0163}$ & $-0.15^{+0.13}_{-0.13}$ & $0.1089^{+0.0130}_{-0.0115}$ & $-0.07^{+0.14}_{-0.14}$ & $0.1460^{+0.0048}_{-0.0047}$ & $-0.68^{+0.60}_{-0.61}$ & $0.1038^{+0.0036}_{-0.0034}$ & $-0.32^{+0.62}_{-0.61}$ \\
rMS & $0.1348^{+0.0049}_{-0.0049}$ & $0.02^{+0.05}_{-0.04}$ & $0.1394^{+0.0053}_{-0.0050}$ & $-0.09^{+0.05}_{-0.05}$ & $0.1369^{+0.0018}_{-0.0018}$ & $0.10^{+0.18}_{-0.19}$ & $0.1301^{+0.0018}_{-0.0017}$ & $-0.38^{+0.19}_{-0.20}$ \\
\hline
\end{tabular}
\caption{Posterior medians and $16-84^\mathrm{th}$ percentile intervals of the equipartition parameters for the radial (R) and tangential (T) components, fitted to the \jwst\ stars in F3.}
\label{tab:bestfit_eq}
\end{table*}

\label{lastpage}
% \bsp
\end{document}